\documentclass{article} 
\usepackage{iclr2027_conference,times}
\iclrfinalcopy  

\usepackage{amsmath,amsfonts,bm}

\def\eqref#1{equation~\ref{#1}}

\def\1{\bm{1}}

\DeclareMathAlphabet{\mathsfit}{\encodingdefault}{\sfdefault}{m}{sl}
\SetMathAlphabet{\mathsfit}{bold}{\encodingdefault}{\sfdefault}{bx}{n}

\usepackage{hyperref}
\usepackage{url}

\usepackage[utf8]{inputenc} 
\usepackage[T1]{fontenc}    
\usepackage{hyperref}       
\usepackage{url}            
\usepackage{booktabs}       
\usepackage{placeins}
\usepackage{enumitem}
\usepackage{amsfonts}       
\usepackage{nicefrac}       
\usepackage{microtype}      
\usepackage{xcolor}         
\usepackage{algorithm}
\usepackage{algpseudocode}
\usepackage{mathtools, amssymb, amsmath, amsfonts}
\usepackage{amsthm}

\usepackage[font=small,skip=4pt]{caption}
\usepackage{subcaption}

\usepackage{tikz}
\usetikzlibrary{arrows.meta,calc}
\usepackage{graphicx}
\graphicspath{{./}}

\title{Adaptive Scenario Generation for KL-DRO Around a Learned Density Model}

\author{%
  Aron Brenner \& Saurabh Amin \\
  Laboratory for Information \& Decision Systems \\
  Massachusetts Institute of Technology \\
  Cambridge, MA 02139, USA \\
  \texttt{\{abrenner,amins\}@mit.edu}
}

\begin{document}

\maketitle
\lhead{}\renewcommand{\headrulewidth}{0pt}  

\begin{abstract}
    Distributionally robust optimization (DRO) hedges data-driven two-stage stochastic programs against error in the distribution estimated from a finite record, and standard ambiguity sets differ in how they treat realizations absent from that record. A divergence ball around the empirical distribution contains only reweightings of the record, whereas a transport ball also contains distributions supported off the record, including on realizations the data-generating process cannot produce, against which the decision is then hedged as well. We instead propose a Kullback--Leibler (KL) ball centered at a density model fitted to the record, the learned nominal, which lets the adversary up-weight costly unobserved realizations in proportion to the probability the model assigns them. Solving the resulting problem poses a second difficulty, because it can accommodate only a few scenarios, and an independent sample of that size is likely to omit the rare realizations on which the worst case concentrates. We therefore build the scenario set adaptively, alternating between solving the finite problem and drawing new realizations from the learned nominal, steered toward the worst case at the current decision, with importance weights that correct the draws back to the model. On a multi-item newsvendor problem and a capacity expansion problem on realistic grid data, centering the ball at the learned nominal rather than at the record lowers cost under distribution shift, at comparable in-distribution cost, when the record is scarce and the model class captures structure that lets it extrapolate beyond the record. The advantage fades as the record grows or as the shift departs from the structure the model has learned.
\end{abstract}

\section{Introduction}\label{sec:intro}

Two-stage stochastic programs model decisions that must be committed before uncertainty is realized, with recourse actions taken after, in applications spanning finance, supply chains, telecommunications, and energy systems \citep{wallace_applications_2005, birge_introduction_2011}. We consider their data-driven, \textit{distributionally robust} form,
\begin{align} \label{eq:dro}
    \min_{x \in \mathcal X} \; \Big\{ c(x) + \max_{\mathbb P \in \mathcal P(\mathbb Q)} \mathcal R_{\mathbb P}\big[ Q(x,\xi) \big] \Big\},
    \qquad
    Q(x,\xi) \coloneqq \min_{y \in \mathcal Y(\xi, x)} q(y),
\end{align}
where $x$ denotes here-and-now decisions with cost $c(x)$, $\xi \in \mathbb R^D$ the random vector, $Q(x,\xi)$ the cost of the wait-and-see decisions $y$, $\mathcal P(\mathbb Q)$ an \emph{ambiguity set} of distributions considered plausible around a nominal distribution $\mathbb Q$ (in the standard setting, the empirical distribution $\hat{\mathbb Q}_N$ of a record of $N$ past realizations), and $\mathcal R_{\mathbb P}$ a risk measure, such as the expectation or the conditional value at risk (CVaR). We are motivated by problems whose cost is dominated by tail events, in which several dimensions of uncertainty move against the plan at once, such as a demand surge that arrives while a supplier is down.

Such problems expose two gaps in how data-driven distributionally robust optimization (DRO) treats realizations absent from the record. The first lies in the ambiguity set (Sec.~\ref{sec:literature}). Divergence balls around the empirical distribution contain only reweightings of the record, while Wasserstein balls and moment-based sets admit unobserved realizations but judge them only by their distance from the record or by a few moments. Each family carries a statistical guarantee tied to the record \citep{duchi2021statistics, mohajerin_esfahani_data-driven_2018}, but none uses how the components of the uncertainty move together in the tail, the structure that distinguishes plausible unobserved realizations from implausible ones. The second gap lies in the finite scenario set on which the problem is actually solved, whose size is limited because each scenario adds a copy of the recourse problem. A divergence ball can only reweight this set. Scenario reduction and problem-driven selection choose it from the record or from a sample drawn once, so the set is likely to miss the rare realizations on which the worst case concentrates.

Both gaps concern realizations that the record lacks, and both could be narrowed by a model that generates plausible ones. High-resolution operational records, often spanning only a few years, are available in many applications of stochastic programming, and generative models trained on such records can capture their dependence structure and respect known properties of the process, including the behavior of its extremes \citep{boulaguiem2022modeling, pandey2025heavy, chang2026eta}. Such a model offers both a new center for the ambiguity set, addressing the first gap, and a new source of scenarios, addressing the second. This raises the question of \emph{how a generative model learned from a short record can be integrated into risk-averse stochastic programming, so that a limited set of scenarios hedges against tail risks the record underrepresents}.

We propose Kullback--Leibler DRO (KL-DRO) centered at a learned nominal: we take $\mathbb Q$ in \eqref{eq:dro} to be a density model $\mathbb Q_\theta$ trained on the record, such as a normalizing flow, a Gaussian copula, or a diffusion model, and the ambiguity set to be
\begin{align} \label{eq:gen_ball}
    \mathcal P_\rho(\mathbb Q_\theta) \coloneqq \big\{ \mathbb P : \mathrm{KL}(\mathbb P \,\|\, \mathbb Q_\theta) \leq \rho \big\}.
\end{align}
Over this ball, the adversary can up-weight costly realizations absent from the record in proportion to the probability $\mathbb Q_\theta$ assigns them, so what it can reach is governed by the learned nominal rather than by a distance from the record, which addresses the first gap. To close the second, we build the scenario set adaptively, alternating between solving the finite problem and drawing new realizations from the learned nominal, steered toward the worst case at the current decision, with importance weights that center the ball at an estimate of $\mathbb Q_\theta$. Centering the ball at a learned nominal does, however, give up the finite-sample guarantees of balls centered at the record, so we choose the radius empirically (Sec.~\ref{sec:experiments}).

\paragraph{Contributions.} Our contributions are as follows.
\begin{itemize}
    \item \textbf{Distributionally robust optimization with a learned nominal.} We pose KL-DRO with the ambiguity set centered at a learned generative model, extending balls centered at parametric models or Bayesian posteriors for single-stage objectives \citep{watson_approximate_2016, dellaporta2025decision} to high-dimensional nominals and two-stage objectives.

    \item \textbf{Finite scenario approximations for tail-dominated problems.} We build the scenario set by adaptive importance sampling toward the ball's worst case at each decision, so that the scenarios are drawn where the adversary places its weight.

    \item \textbf{Evaluations in controlled and realistic settings.} We evaluate the framework on a multi-item newsvendor problem with a known distribution and controlled record size and shift, and on a capacity expansion problem over a 20-zone network with realistic demand and renewable data and an out-of-distribution test year.
\end{itemize}

\section{Background} \label{sec:background}

\paragraph{Risk-averse stochastic programming.}\label{sec:setting} Given a limited record of past realizations, we seek decisions that are cost-effective both \emph{in distribution}, even when the record is too small to learn the data-generating process, and \emph{out of distribution}, when that process changes at evaluation time. To weigh both typical and tail costs, we take $\mathcal R$ to be the mean--CVaR objective, which can be defined as the minimum over a scalar $\zeta$ of an expectation \citep{rockafellar2000optimization},
\begin{align}
    \mathcal R_{\mathbb P}[Q(x,\xi)] &= (1-\beta)\,\mathbb E_{\mathbb P}[Q(x,\xi)] + \beta\,\mathrm{CVaR}_{\alpha,\mathbb P}[Q(x,\xi)] \nonumber \\
    &= \min_{\zeta \in \mathbb R} \big\{ \zeta + \mathbb E_{\mathbb P}[\psi_\zeta(Q(x,\xi))] \big\}, \label{eq:mean_cvar} \\
    \psi_\zeta(u) &= (1-\beta)(u - \zeta) + \tfrac{\beta}{1-\alpha}\,(u - \zeta)_+, \nonumber
\end{align}
where $\alpha$ is the tail level, $\beta \in [0, 1]$ the weight on the tail, and the minimizing $\zeta$ the value at risk at level $\alpha$. Throughout, we assume that the recourse problem is feasible (relatively complete recourse) and bounded for every $x \in \mathcal X$ and every $\xi$ in the support of the nominal, so that $Q(x,\xi)$ is finite.

\paragraph{Kullback--Leibler DRO.}\label{sec:kldro} For a nominal distribution $\mathbb Q$, empirical or otherwise, the Kullback--Leibler (KL) ambiguity set $\{\mathbb P : \mathrm{KL}(\mathbb P \,\|\, \mathbb Q) \leq \rho\}$ is a standard choice for two-stage problems \citep{ben-tal_robust_2013, bayraksan_data-driven_2015}. Because every member assigns zero probability wherever $\mathbb Q$ does, the ball reweights the realizations $\mathbb Q$ supports and adds none, and when centered at the empirical record, it changes only the probabilities of the recorded scenarios. The ball is computationally attractive because dualizing the KL constraint reformulates the inner maximization of \eqref{eq:dro} as a minimization (\citealp{hu2013kullback}; App.~\ref{app:optimization}),
\begin{align}
    \max_{\mathbb P \,:\, \mathrm{KL}(\mathbb P \,\|\, \mathbb Q) \leq \rho} \mathcal R_{\mathbb P}[Q(x,\xi)]
    = \min_{\zeta,\, \lambda > 0} \Big\{ \zeta + \lambda \rho + \lambda \log \mathbb E_{\mathbb Q}\big[ \exp\big(\psi_\zeta(Q(x,\xi))/\lambda\big) \big] \Big\}. \label{eq:kl_dual}
\end{align}
For $\rho > 0$, both sides are finite only if $\mathbb E_{\mathbb Q}[\exp(Q(x,\xi)/\lambda)] < \infty$ for some $\lambda > 0$, which holds for any finite nominal and any bounded cost. The dual is jointly convex in $\zeta$ and $\lambda$, and in $x$ as well when the recourse cost is convex in $x$, as it is for linear recourse. In particular, for a finite nominal $\hat{\mathbb Q} = \sum_{i=1}^K w_i\, \delta_{\xi_i}$, a set of $K$ scenarios with weights $w_i > 0$ summing to one, the expectation is a weighted sum, and \eqref{eq:dro} over the ball becomes, when $\mathcal X$ and $c$ are convex, the single convex program
\begin{align} \label{eq:master}
    \min_{x \in \mathcal X,\, \zeta,\, \lambda > 0} \; c(x) + \zeta + \lambda \rho + \lambda \log \sum_{i=1}^K w_i \exp\big(\psi_\zeta(Q(x,\xi_i))/\lambda\big),
\end{align}
in which each scenario contributes one copy of the recourse problem and the risk measure adds one variable. The empirical measure $\hat{\mathbb Q}_N$ is the special case with $K = N$ and $w_i = 1/N$. Since the reformulation holds for any finite nominal, two modeling choices remain open: which realizations the adversary can reach, which is set by $\mathbb Q$, and which finite $\hat{\mathbb Q}$ the problem is solved on. We discuss them in turn.

\paragraph{Ambiguity centered at a learned nominal.}\label{sec:learned_nominal} The first choice determines which realizations the adversary can reach. Ideally, the nominal places mass on realizations the record lacks but not on those the data-generating process cannot produce, so that the decision is hedged against unrecorded events without paying for implausible ones. A density model $\mathbb Q_\theta$ fitted to the record places mass between and beyond the recorded realizations, distributed according to the dependence structure it has learned, so the ball around it contains distributions that weight events the record lacks. How well it withholds mass from implausible realizations depends on how faithfully it captures the dependence structure of the process. Because $\mathbb Q_\theta$ is itself an estimate, the ball around it also hedges against error in the model, as in \citet{watson_approximate_2016} and \citet{dellaporta2025decision}. We therefore adopt the ambiguity set \eqref{eq:gen_ball}, the KL ball above centered at $\mathbb Q_\theta$ rather than at the record. To make the nominal easy to sample and its draws easy to weight, we take $\mathbb Q_\theta$ to be the pushforward of the standard normal distribution $\varphi \coloneqq \mathcal N(0, I)$, whose density we also write $\varphi$, by a deterministic decoder $G_\theta$, so that sampling amounts to drawing $z \sim \varphi$ and applying $G_\theta$. Normalizing flows, Gaussian copulas, and diffusion models with deterministic samplers are of this form, and the known latent density is what gives the sampler of Sec.~\ref{sec:methodology} exact importance weights.

\paragraph{Finite scenario approximation.}\label{sec:finite} The second choice is the finite nominal on which the problem is solved. The finite program \eqref{eq:master}, which we call the master problem, grows with each scenario, so the number of scenarios $K$ must be small, and the decision is then largely determined by which $K$ scenarios the master sees. When the realizations that drive the decision are rare, such as the few periods in which a demand surge coincides with a supplier outage, $K$ independent draws are likely to miss them. Scenario reduction and problem-driven selection mitigate this by choosing the $K$ scenarios from a larger pool (Sec.~\ref{sec:literature}), but a pool fixed in advance cannot supply realizations it lacks. A learned nominal, by contrast, can be sampled wherever it has mass, so the $K$ scenarios can be generated where the decision is sensitive. Because such scenarios are not drawn from $\mathbb Q_\theta$ itself, each carries a base weight $w_i$, an importance weight back to $\mathbb Q_\theta$, and we solve \eqref{eq:master} with these weights normalized to sum to one, so that the ball is centered at an importance-weighted estimate of $\mathbb Q_\theta$ rather than at the distribution the scenarios were drawn from.

\section{Related Work} \label{sec:literature}

\paragraph{Ambiguity sets.}
Divergence-based sets \citep{ben-tal_robust_2013, bayraksan_data-driven_2015} reweight the empirical record and so preclude unobserved realizations, or, with the arguments of the divergence reversed, admit them only on a fixed finite support \citep{vanparys2021data}. Transport-based sets \citep{mohajerin_esfahani_data-driven_2018, gao_distributionally_2023} admit them, but judge them by a cost that measures how far a recorded point moves and says nothing directly about whether the moved point is plausible. The Sinkhorn variant smooths this cost against a reference measure chosen in advance, such as a Gaussian, which sets where the worst case can place mass \citep{wang_sinkhorn_2021}. Moment-based sets \citep{delage2010distributionally} constrain only low-order statistics. Closest in motivation are works that learn part of the set: the transport cost, from labels or the data manifold \citep{blanchet_data-driven_2019, liu2022distributionally}, or the nominal itself, a fitted model or Bayesian posterior at which a KL-based set is centered \citep{watson_approximate_2016, dellaporta2025decision}.

\paragraph{Scenario construction.}
Distribution-driven methods match specified moments \citep{hoyland_heuristic_nodate} or reduce a pool under a probability metric \citep{dupacova_scenario_2003, heitsch_scenario_2003} without reference to the decision. Problem-driven methods cluster by decision-based dissimilarity or minimize implementation error \citep{hewitt_decision-based_2022, keutchayan_problem-driven_2023}, or, for tail risk measures, sample only the region of the support on which the risk measure depends \citep{fairbrother_problem-driven_2022}. All of these methods construct the set once, before the problem is solved, whereas the importance sampling methods of Sec.~\ref{sec:methodology} adapt the proposal to the decision.

\paragraph{Generative models in optimization.}
A more recent line of work samples scenarios from a generative model fitted to the record \citep{chen2018model, jiang2018scenario} or trains the generator on the downstream decision loss \citep{zhao_diffusion-dfl_2025, islip2025contextualscenariogenerationtwostage}, producing scenarios but no ambiguity set. A second line makes the generator the adversary, through a learned transport map to a continuous least favorable distribution \citep{xu_flow-based_2024, cheng_worst-case_2025}, a family of parametric generative models within a KL budget \citep{michel2021modeling}, a family of generators of bounded reconstruction loss \citep{wen_distributionally_2026}, a Sinkhorn comparison between conditional generators \citep{zhang2026generative}, or, for two-stage robust optimization, a search of a generator's latent space for a single worst-case realization \citep{brenner2025deep}. In each, the worst case is obtained by optimizing over that parameterization, and all but the last are posed for single-stage objectives. We instead center a divergence ball at the generator, which leaves the generator as trained, gives the worst case in closed form, and keeps the likelihood ratios exact.

\section{Adaptive Importance Sampling Algorithm} \label{sec:methodology}

\begin{figure}[t]
    \centering
    \resizebox{0.82\linewidth}{!}{\input{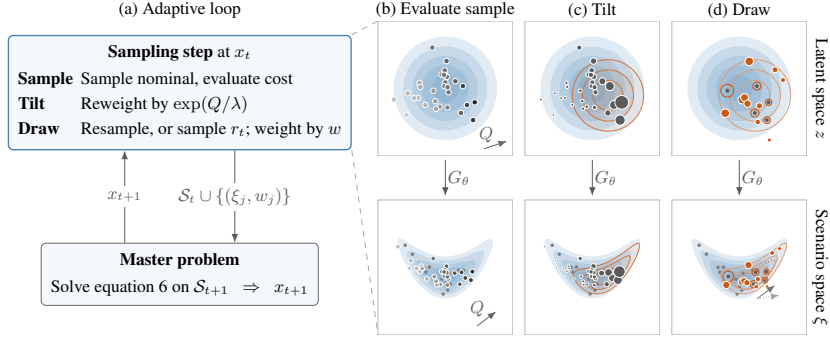}%
\providecommand{\figfont}{\rmfamily}%
\providecommand{\figdenscol}{70,130,180}     %
\providecommand{\figpropcol}{205,90,20}     %
\providecommand{\figdrawcol}{205,90,20}      %
\providecommand{\figcostlow}{205,205,205}    %
\providecommand{\figcosthigh}{35,35,35}     %
\providecommand{\figpilotcol}{90,90,90}   %
\providecommand{\figframecol}{175,180,190}%
\providecommand{\figboxcol}{244,247,252}%
\providecommand{\figsampborder}{70,130,180}    %
\providecommand{\densstyle}{fill=denscol, opacity=0.16}%
\providecommand{\propstyle}{draw=propcol!85, line width=0.45pt, fill=propcol, fill opacity=0.05}%
\providecommand{\propoutline}{draw=propcol, line width=0.7pt}%
\providecommand{\framestyle}{draw=framecol, line width=0.4pt}%
\providecommand{\markeredge}{draw=white, line width=0.25pt}%
\providecommand{\minimalbox}{1}             %
\definecolor{denscol}{RGB}{\figdenscol}%
\definecolor{propcol}{RGB}{\figpropcol}%
\definecolor{drawcol}{RGB}{\figdrawcol}%
\definecolor{costlow}{RGB}{\figcostlow}%
\definecolor{costhigh}{RGB}{\figcosthigh}%
\definecolor{pilotcol}{RGB}{\figpilotcol}%
\definecolor{framecol}{RGB}{\figframecol}%
\definecolor{boxcol}{RGB}{\figboxcol}%
\definecolor{sampborder}{RGB}{\figsampborder}%
\def\colx{4.95}\def\gap{0.22}\def\rowtop{3.55}\def\rowbot{0.30}\providecommand{\scenshift}{0.35}\providecommand{\figndraw}{10}%
\pgfmathsetmacro{\colxend}{\colx+3*\panelside+2*\gap+0.45}%
\begin{tikzpicture}[x=1cm, y=1cm, font=\figfont\footnotesize,
  wbox/.style={draw=black!60, fill=boxcol, rounded corners=2.5pt, inner sep=5pt, align=center, font=\figfont\footnotesize},
  arr/.style={-{Latex[length=1.6mm]}, black!65, line width=0.5pt},
  alab/.style={midway, fill=white, inner sep=2.5pt, font=\figfont\footnotesize, align=center},
  panel/.style/.expand once=\framestyle,
  medge/.style/.expand once=\markeredge,
  ptitle/.style={anchor=south, inner sep=1.5pt, font=\figfont\footnotesize}]

\begin{scope}
  \node[wbox, draw=sampborder, line width=0.7pt] (samp) at (2.35,4.7) {%
    {\footnotesize\textbf{Sampling step} at $x_t$}\\[4pt]
    \begin{tabular}{@{}l@{\hspace{4pt}}l@{}}
      \textbf{Sample} & Sample nominal, evaluate cost \\[3pt]
      \textbf{Tilt}  & Reweight by $\exp(Q/\lambda)$ \\[3pt]
      \textbf{Draw}  & Resample, or sample $r_t$; weight by $w$
    \end{tabular}};
  \node[wbox] (master) at (2.35,1.4) {%
    {\footnotesize\textbf{Master problem}}\\[4pt]
    Solve \eqref{eq:master_round} on $\mathcal S_{t+1}$ \; $\Rightarrow$ \; $x_{t+1}$};
  \draw[arr] ([xshift=1.0cm]samp.south) -- node[alab] {$\mathcal S_t \cup \{(\xi_j, w_j)\}$} ([xshift=1.0cm]master.north);
  \draw[arr] ([xshift=-1.0cm]master.north) -- node[alab] {$x_{t+1}$} ([xshift=-1.0cm]samp.south);
  \node[ptitle] at (samp.north |- 0,{\rowtop+\panelside}) {(a) Adaptive loop};
\end{scope}

\def\colx{5.95}
\def\gap{0.22}
\def\rowtop{3.55}

\draw[dashed, black!35, line width=0.4pt] (samp.north east) -- (\colx, \rowtop+\panelside);
\draw[dashed, black!35, line width=0.4pt] (samp.south east) -- (\colx, \rowbot);

\foreach \col/\ttl in {0/{(b) Evaluate sample}, 1/{(c) Tilt}, 2/{(d) Draw}} {
  \pgfmathsetmacro{\px}{\colx + \col*(\panelside+\gap)}
  \node[ptitle] at (\px+0.5*\panelside,\rowtop+\panelside) {\ttl};
  \draw[arr] (\px+0.5*\panelside, \rowtop-0.07) -- node[right, font=\figfont\footnotesize, inner sep=1pt] {$G_\theta$} (\px+0.5*\panelside, \rowbot+\panelside+0.07);
}

\begin{scope}[shift={(\colx,\rowtop)}]
  \clip (0,0) rectangle (\panelside,\panelside);
  \expandafter\priorlatent\expandafter{\densstyle}
  \draw[arr, black!60] (1.930,0.130) -- node[midway, above left, inner sep=1pt, font=\figfont\footnotesize] {$Q$} (2.348,0.268);
  \foreach \x/\y/\c in \pilotlatentpts { \path[fill=costhigh!\c!costlow, medge] (\x,\y) circle (1.4pt); }
  \path[panel] (0,0) rectangle (\panelside,\panelside);
\end{scope}
\begin{scope}[shift={(\colx,\rowbot)}]
  \clip (0,0) rectangle (\panelside,\panelside);
  \begin{scope}[yshift=\scenshift cm]
  \expandafter\priorscen\expandafter{\densstyle}
  \foreach \x/\y in \recordpts { \fill[black!50] (\x,\y) circle (0.9pt); }
  \foreach \x/\y/\c in \pilotscenpts { \path[fill=costhigh!\c!costlow, medge] (\x,\y) circle (1.25pt); }
  \end{scope}
  \draw[arr, black!60] (1.800,0.160) -- node[midway, above left, inner sep=1pt, font=\figfont\footnotesize] {$Q$} (2.152,0.424);
  \path[panel] (0,0) rectangle (\panelside,\panelside);
\end{scope}

\pgfmathsetmacro{\pxc}{\colx + \panelside + \gap}
\begin{scope}[shift={(\pxc,\rowtop)}]
  \clip (0,0) rectangle (\panelside,\panelside);
  \expandafter\priorlatent\expandafter{\densstyle}
  \expandafter\proplatent\expandafter{\propstyle}
  \foreach \x/\y/\r in \tiltlatentpts { \path[fill=pilotcol, medge] (\x,\y) circle (\r pt); }
  \path[panel] (0,0) rectangle (\panelside,\panelside);
\end{scope}
\begin{scope}[shift={(\pxc,\rowbot)}]
  \clip (0,0) rectangle (\panelside,\panelside);
  \begin{scope}[yshift=\scenshift cm]
  \expandafter\priorscen\expandafter{\densstyle}
  \expandafter\propscen\expandafter{\propstyle}
  \foreach \x/\y in \recordpts { \fill[black!50] (\x,\y) circle (0.9pt); }
  \foreach \x/\y/\r in \tiltscenpts { \path[fill=pilotcol, medge] (\x,\y) circle (0.85*\r pt); }
  \end{scope}
  \path[panel] (0,0) rectangle (\panelside,\panelside);
\end{scope}

\pgfmathsetmacro{\pxd}{\colx + 2*(\panelside + \gap)}
\begin{scope}[shift={(\pxd,\rowtop)}]
  \clip (0,0) rectangle (\panelside,\panelside);
  \expandafter\priorlatent\expandafter{\densstyle}
  \proplatent{draw=propcol!60, line width=0.4pt}
  \foreach \x/\y/\r in \sellatentpts { \fill[pilotcol] (\x,\y) circle (1.0pt); \draw[drawcol, line width=0.6pt] (\x,\y) circle (\r pt); }
  \foreach \x/\y/\r [count=\k] in \drawlatentpts { \ifnum\k>\figndraw\relax\else\path[fill=drawcol, medge] (\x,\y) circle (\r pt);\fi }
  \path[panel] (0,0) rectangle (\panelside,\panelside);
\end{scope}
\begin{scope}[shift={(\pxd,\rowbot)}]
  \clip (0,0) rectangle (\panelside,\panelside);
  \begin{scope}[yshift=\scenshift cm]
  \expandafter\priorscen\expandafter{\densstyle}
  \foreach \x/\y in \recordpts { \fill[black!50] (\x,\y) circle (0.9pt); }
  \propscen{draw=propcol!60, line width=0.4pt}
  \foreach \x/\y/\r in \selscenpts { \fill[pilotcol] (\x,\y) circle (0.9pt); \draw[drawcol, line width=0.6pt] (\x,\y) circle (0.85*\r pt); }
  \foreach \x/\y/\r [count=\k] in \drawscenpts { \ifnum\k>\figndraw\relax\else\path[fill=drawcol, medge] (\x,\y) circle (0.85*\r pt);\fi }
  \draw[arr, black!55] \xgradA -- \xgradB;
  \draw[arr, black!35, densely dotted] \xgradA -- \xgradC;
  \expandafter\path\expandafter[\propoutline, propcol!75, densely dotted] plot[smooth cycle] coordinates {\propimgtwo};
  \end{scope}
  \path[panel] (0,0) rectangle (\panelside,\panelside);
\end{scope}

\node[anchor=south, rotate=-90, font=\figfont\footnotesize, inner sep=1pt] at (\colx+3*\panelside+2*\gap+0.06, \rowtop+0.5*\panelside) {Latent space $z$};
\node[anchor=south, rotate=-90, font=\figfont\footnotesize, inner sep=1pt] at (\colx+3*\panelside+2*\gap+0.06, \rowbot+0.5*\panelside) {Scenario space $\xi$};
\end{tikzpicture}}
    \caption{\textbf{(a)} The adaptive loop. \textbf{(b--d)} One sampling step in a 2-D slice of the latent space (contours of $\varphi$; cost rises to the upper right): (b) an initial sample is evaluated; (c) it is tilted by $\exp(Q/\lambda)$, and a Gaussian proposal $g_t$ (orange) is fitted; (d) new scenarios are resampled (rings) or drawn from the proposal (filled), with weights back to the nominal (marker area). \textbf{Bottom row.} The same steps after decoding by $G_\theta$: new scenarios reach beyond the record (gray), and the target moves with the decision (dotted).}
    \label{fig:overview}
\end{figure}

Section~\ref{sec:finite} called for a finite scenario set generated where the decision is sensitive and weighted back to $\mathbb Q_\theta$. We propose building it iteratively by adaptive importance sampling toward a target known up to a normalizing constant \citep{bugallo2017adaptive}, re-targeting as the decision changes, as \citet{dantzig1990parallel}, \citet{infanger_monte_1992}, \citet{parpas_importance_2015}, and \citet{aolaritei_stochastic_2026} do for a known density. Each round tilts a sample of the model toward the worst case at the current decision and forms the batch either by resampling that sample, which needs only draws of the model, or by drawing from a Gaussian fitted to the tilt by weighted moment matching, a cross-entropy step \citep{rubinstein2004cross} that can reach realizations beyond the initial sample. Our approach differs from these methods in its target and its nominal. First, our target is the worst case of the KL ball, so the draws supply the realizations the adversary up-weights (Sec.~\ref{sec:tilt}), whereas \citet{aolaritei_stochastic_2026} target the variance of a stochastic gradient. Second, we draw from a learned density model (Sec.~\ref{sec:etas}), whereas the methods above assume a known density and \citet{parpas_importance_2015} fit a kernel density estimate to Markov chain samples, which degrades in high dimension. \citet{kim_generating_nodate} also steer a generative model toward costly inputs with importance weights, but to train a predictor rather than to make a decision. Figure~\ref{fig:overview} summarizes the resulting loop, whose components the rest of this section develops: the master solved each round (Sec.~\ref{sec:loop}), the target toward which scenarios are drawn (Sec.~\ref{sec:tilt}), and the two ways a batch is drawn (Sec.~\ref{sec:etas}).

\subsection{Alternating Master and Sampler} \label{sec:loop}

\begin{algorithm}[!b]
\caption{Adaptive scenario acquisition.}
\label{alg:etas}
\footnotesize
\begin{algorithmic}[1]
\Require Nominal $\mathbb Q$ that can be sampled; radii $\rho$ and $\rho_{\mathrm{ref}}$; initial sample size $M$; batch size $n$; rounds $T$; re-targeting rounds $\mathcal T \ni 0$; mixture weight $a$
\State Choose an initial decision $x_0$; set $\mathcal S_0 \gets \emptyset$
\For{$t = 0, \dots, T-1$}
    \If{$t \in \mathcal T$} \Comment{Re-target; otherwise keep the last}
        \State Draw $\xi_1, \dots, \xi_M \sim \mathbb Q$ and evaluate $Q_i \gets Q(x_t, \xi_i)$ \Comment{Initial sample}
        \State Bisect $\lambda_t$ so that $\tau_i \propto \exp(Q_i/\lambda_t)$ satisfies $\sum_i \tau_i \log (M\tau_i) = \rho_{\mathrm{ref}}$ \Comment{Tilt}
    \EndIf
    \If{resampling}
        \State Resample $n$ indices $\propto \tau$; $\mathcal B_t \gets \{(\xi_i, c_i/(M\tau_i))\}$ over distinct $i$ \Comment{Resampling}
    \Else
        \State Fit $g_t$ to the tilted latent points of the initial sample \Comment{Fitted proposal}
        \State Draw from $r_t = (1-a)\,\varphi + a\,g_t$ with weights $\varphi/r_t$; keep $n$ draws as $\mathcal B_t$, or resample $n$ by the tilt as above
    \EndIf
    \State $\mathcal S_{t+1} \gets \mathcal S_t \cup \mathcal B_t$
    \State Solve \eqref{eq:master_round} on $\mathcal S_{t+1}$ for $x_{t+1}$ \Comment{Master}
\EndFor
\State \Return $x_T$
\end{algorithmic}
\end{algorithm}

\paragraph{The master on the scenario set.} Each round draws a batch of $n$ scenarios at the incumbent $x_t$ with base weights back to the nominal (Sec.~\ref{sec:etas}), adds it to the set held, and solves \eqref{eq:master} on that set for $x_{t+1}$ (Algorithm~\ref{alg:etas}). Scenarios drawn at earlier decisions can be kept at later ones, because a batch's weights depend only on how it was drawn. Hence, for any decision $x$ fixed in advance, the weighted sum of the batch's costs, before normalization, estimates $n$ times the nominal expected cost without bias. Since \eqref{eq:master} requires weights that sum to one, the weights of all rounds are normalized together, $\bar w_i = w_i / \sum_j w_j$ over the $K_{t+1}$ scenarios held. The master's objective is therefore a biased estimate of its counterpart under $\mathbb Q_\theta$, through this division by a random total and through the logarithm and the minimizations over $\zeta$ and $\lambda$ below. With these weights, the master on the held scenarios is
\begin{align} \label{eq:master_round}
    \min_{x \in \mathcal X,\, \zeta,\, \lambda > 0} \; c(x) + \zeta + \lambda\rho + \lambda \log \sum_{i=1}^{K_{t+1}} \bar w_i \exp\big(\psi_\zeta(Q(x,\xi_i))/\lambda\big).
\end{align}
At a decision $x$, the adversary's worst case is available in closed form, $p_i \propto \bar w_i \exp(\psi_{\zeta^*}(Q(x,\xi_i))/\lambda^*)$, where $\zeta^*$ and $\lambda^*$ attain the minimum at $x$, so that $\mathrm{KL}(p \,\|\, \bar w) = \rho$. When \eqref{eq:master_round} is too large to solve jointly, as in the capacity expansion study, this closed form offers a cheaper alternative, in which the master is solved with the weights fixed at the worst case computed at $x_t$, which requires $K_t$ recourse evaluations. Weak duality then bounds the resulting optimality gap on the held scenarios at the cost of one more master solve (App.~\ref{app:optimization}).

\paragraph{Sampler and radii.} The sampler, which chooses the scenarios that enter the set, tilts an initial sample of $M$ draws from the nominal toward the worst case at a reference radius $\rho_{\mathrm{ref}}$, $\tau_i \propto \exp(Q(x_t, \xi_i)/\lambda_t)$ with $\sum_i \tau_i \log(M\tau_i) = \rho_{\mathrm{ref}}$, and forms the batch from this tilt (Sec.~\ref{sec:etas}). The two radii thus play different roles, $\rho_{\mathrm{ref}}$ determining which scenarios enter the set and $\rho$ how the master weights them, so that $\rho = 0$ with $\rho_{\mathrm{ref}} > 0$ solves the nominal problem with a decision-aware sampler. We take $\rho_{\mathrm{ref}} \geq \rho$ as a heuristic, because a sampler that reaches beyond the adversary only costs efficiency, whereas one that reaches less leaves tail scenarios out of the set, and no reweighting can restore them. The tilt is the worst case under the expectation rather than under the mean--CVaR objective, but because $\psi_\zeta$ is nondecreasing, both worst cases rank scenarios by cost in the same way. Since the weights correct any proposal back to the nominal, this choice affects efficiency but not the quantity being estimated.

\paragraph{Cost of a round.} Each initial sample costs $M$ recourse evaluations, which run in parallel, whereas only the batch of $n$ enters the master, whose size grows with each scenario. To save evaluations, the initial sample need only be redrawn and re-tilted in the rounds $t \in \mathcal T$, because the tilt depends on the decision only through the costs at $x_t$ and so changes little once the incumbent settles. Later batches reuse the last initial sample and proposal, and their weights remain exact, although resampling the same initial sample again does not reduce its own sampling error.

\subsection{The Sampling Target} \label{sec:tilt}

\paragraph{Latent space.} The sampler works in the latent space, where the nominal is the standard Gaussian $\varphi$. A simple parametric family would match $\mathbb Q_\theta$ poorly in the ambient space, but because $\varphi$ is known, a proposal from such a family has likelihood ratios in closed form in the latent space, without evaluating the density of $\mathbb Q_\theta$. These ratios correct the expectation of any function of $z$, including the cost $Q(x, G_\theta(z))$ of the decoded scenario. The whole problem can also be posed in the latent space,
\begin{align} \label{eq:latent_problem}
    \min_{x \in \mathcal X} \; \Big\{ c(x) + \max_{p \,:\, \mathrm{KL}(p \,\|\, \varphi) \leq \rho} \; \mathcal R_{z \sim p}\big[ Q(x, G_\theta(z)) \big] \Big\},
\end{align}
in which the adversary reweights $\varphi$. This latent problem is equivalent to the original one, because decoding maps the latent ball onto the ball around $\mathbb Q_\theta$ for any measurable $G_\theta$. Pushing two distributions through the same map cannot increase the divergence between them \citep{cover2006elements}, so the latent ball decodes into the ambient ball. Conversely, any $\mathbb P$ in the ambient ball is the image of $\varphi$ reweighted by $\frac{d\mathbb P}{d\mathbb Q_\theta} \circ G_\theta$, whose divergence from $\varphi$ equals $\mathrm{KL}(\mathbb P \,\|\, \mathbb Q_\theta)$. For a law-invariant $\mathcal R$, such as the expectation and CVaR \citep{kuhn2025distributionally}, the two problems therefore have the same value and minimizers, whether or not $G_\theta$ is invertible.

\paragraph{The tilt as a proposal.} Under the expectation, the worst case of \eqref{eq:latent_problem} at a fixed decision is the exponential tilt
\begin{align} \label{eq:tilt}
    \pi(z \mid x) \;\propto\; \varphi(z)\, \exp\big(Q(x, G_\theta(z))/\lambda\big),
\end{align}
with $\lambda$ fixed by the radius. Used as a sampling target, the tilt weights a realization by the level of its cost, whereas the decision responds only to the part of the cost that $x$ can change. The level is therefore a faithful proxy when the costly realizations are those the decision can mitigate, such as unserved load that added capacity would avert, but it fails when the costliest realizations are costly whatever the decision, since the tilt then concentrates the sample where the master cannot respond. The tilt also links the two roles the scenario set plays. Read as robustness, $\pi(\cdot \mid x)$ is the least favorable distribution, and read as estimation, it is the exponential tilting of rare-event sampling \citep{blanchet2011rare}. For large $\lambda$, which corresponds to a small radius, it is close to a mixture of $\varphi$ and the minimum-variance proposal for the expected cost, which reweights $\varphi$ by the cost itself when the cost is nonnegative, and for small $\lambda$, it concentrates on the costliest scenarios, where the master's worst case places its weight. The sampler draws from a proposal that approximates the tilt, and importance weights correct for the approximation, however rough it is. For any proposal $g$ positive wherever $\varphi$ is, draws $z_j \sim g$ with the weight
\begin{align} \label{eq:weight}
    w(z) = \varphi(z) \big/ g(z)
\end{align}
give an unbiased estimate $\tfrac1n \sum_j w(z_j)\, Q(x, G_\theta(z_j))$ of $\mathbb E_\varphi[Q(x, G_\theta(z))]$.

\subsection{Drawing the Batch} \label{sec:etas}

Sampling the target \eqref{eq:tilt} directly is expensive, because it is known only through evaluations of $Q$, each a second-stage solve. Of the existing alternatives, guided sampling of the generator yields draws without likelihood ratios, and Markov chain Monte Carlo \citep{parpas_importance_2015} needs many sequential evaluations of $Q$ per retained sample. We instead form the batch from the tilt $\tau$ of the initial sample, in one of two ways.

\paragraph{Resampling.} The simpler variant resamples $n$ indices with probabilities $\tau$, and a scenario drawn $c_i$ times enters with base weight $w_i = c_i / (M \tau_i)$, which corrects the selection back to the nominal. The batch then holds only realizations in the initial sample, so its reach is governed by $M$. In return, resampling needs only draws, so it also applies to models available only as samplers and to the empirical measure.

\paragraph{Fitted proposal.} To reach beyond the initial sample, the second variant fits a Gaussian $g_t = \mathcal N(\hat\mu, \hat\Sigma)$ to the latent points $z_i$ of the initial sample, $\xi_i = G_\theta(z_i)$, weighted by the tilt, $\hat\mu = \sum_i \tau_i z_i$ and $\hat\Sigma = \sum_i \tau_i (z_i - \hat\mu)(z_i - \hat\mu)^\top$. When the $M$ evaluations are few relative to the latent dimension, $\hat\Sigma$ can be regularized, for example by shrinkage toward its diagonal or, at the extreme, by fixing $\hat\Sigma = I$ so that only the mean shifts. The sampler then draws from the defensive mixture $r_t = (1-a)\,\varphi + a\,g_t$ with $a \in (0, 1)$, decodes $\xi = G_\theta(z)$, and attaches the weight
\begin{align} \label{eq:etas_weight}
    w(z) = \frac{\varphi(z)}{r_t(z)} = \Big[ (1 - a) + a\, \frac{g_t(z)}{\varphi(z)} \Big]^{-1} \leq \frac{1}{1 - a},
\end{align}
with $g_t/\varphi$ in closed form. The weight is exact however $g_t$ was fitted, and since the component $\varphi$ bounds it, a poor fit lowers efficiency but cannot produce an unbounded weight. The draws either enter the master directly with the weights \eqref{eq:etas_weight} or form a pool from which $n$ are resampled by the tilt as above. In that case, the uniform weights $1/M$ are replaced by the weights of \eqref{eq:etas_weight} normalized within the pool, which makes the batch's weighted sum only approximately unbiased (App.~\ref{app:cep_method}).

\paragraph{Choosing a variant.} Resampling needs only draws of the nominal, but its weights vary with the tilt, so a concentrated tilt and a small $n$ give a noisy correction. The fitted proposal, in contrast, reaches beyond the initial sample, with every weight below $1/(1-a)$ when its draws enter the master directly, but it needs latent coordinates and a proposal family that fits the tilt, a fit that degrades with dimension. It should therefore help most when the events that drive the worst case are rare in an initial sample of size $M$ and the tilted target is close to a single Gaussian in the latent space, and least when decoded draws cannot leave the record's range or the map from $z$ to a scenario is strongly nonlinear (Sec.~\ref{sec:newsvendor}).

\section{Experiments} \label{sec:experiments}

We evaluate the framework in two settings. The first, a multi-item newsvendor problem and a standard testbed for DRO \citep{scarf1958, hanasusanto2015multiitem} and data-driven optimization \citep{ban2019big}, has a known demand distribution, so optimality gaps can be computed accurately, and a record size and shift under our control, which lets us isolate the regimes in which a learned nominal improves on the record. The second, a capacity expansion problem for the Texas grid on realistic data, tests whether these findings carry over to a planning problem of practical scale.

\paragraph{Methods.} All methods solve the master \eqref{eq:master} on a finite scenario set and differ in where the scenarios come from, how they are chosen, and how the master hedges. Our method draws scenarios from a learned nominal, either by resampling its draws (KL-Gen) or from a proposal fitted to the tilt (KL-Gen-P), chooses them by the tilt at the current decision, and hedges with a KL ball. We group the comparators by which of these components they change. \emph{Source:} KL-DRO centered at the record (KL-Rec) removes generation; KL-DRO over record points moved toward higher cost by optimal transport (KL-OT) extrapolates by distance from the record instead of by a model; alternative nominals isolate the model class; and in the newsvendor an oracle applies our sampler to the true distribution. \emph{Selection:} cost-space scenario clustering \citep[CSSC;][]{keutchayan_problem-driven_2023} chooses scenarios once, by their effect on the decision. \emph{Hedge:} sample average approximation (SAA) applies none, and Wasserstein DRO \citep[W-DRO;][]{mohajerin_esfahani_data-driven_2018} hedges with a transport ball around the record. Both studies use the mean--CVaR objective with $\beta = 0.25$ and $\alpha = 0.9$, and Tables~\ref{tab:nv_methods} and~\ref{tab:cep_methods} give the implementations. Hedging can lower out-of-distribution cost at the expense of in-distribution cost, so a comparison at a single setting would confound how much a method hedges with how well it hedges. We therefore compare the frontiers of out-of-distribution (OOD) against in-distribution (ID) cost that each method's settings trace.

\subsection{Multi-Item Newsvendor with Substitution} \label{sec:newsvendor}

\paragraph{Problem.}\looseness=-1  A retailer orders five products at a cost of 1 per unit before demand is observed. As recourse, unmet demand for a product can be served from the next product's stock at a penalty of 0.5 per unit (substitution) or lost at a cost of 100 per unit, and leftover stock is salvaged at 0.25. For a single product, the order that minimizes expected cost is the quantile of demand at the critical ratio $(100 - 1)/(100 - 0.25) \approx 0.99$, and the mean--CVaR objective used here raises it further, so the orders lie in the upper tail of demand, which a short record samples poorly. Log demand follows a known Gaussian mixture with a rare shock that raises demand for every product at once (App.~\ref{app:newsvendor}). A decision-maker sees a record of $N \in \{16, 32, 64, 128\}$ periods and is scored under one of four shifted distributions: a level shift in every product's demand, a tail shift to a heavier-tailed shock, a support shift that adds a shock on different products, and a coincidence shift, in which the record's shocks strike one product at a time and the test distribution's strike all products together, with unchanged marginals.

\paragraph{Setup.}\looseness=-1  For each record, we fit five learned nominals: a Gaussian copula; a log-normal; a two-component log-normal mixture, the family of the base distribution; a RealNVP flow; and a Gaussian fitted to raw demand, which captures dependence only through its covariance. KL-Gen-P, KL-Gen, and the oracle run a single round ($T = 1$) from the risk-neutral decision on the record. Because both the distribution the records are drawn from and the shifted distributions are known, we report each cost as a percentage above the optimum under the distribution it is scored on, and Table~\ref{tab:nv-main} reads the resulting frontiers in two ways: (a) the OOD cost within an ID budget of 5 points over SAA, chosen on held-out records without the shifted distribution, and (b) the ID cost needed to halve SAA's OOD cost, which prices protection but is not a deployable rule.

\begin{table}[t]
\centering
\scriptsize
\caption{Newsvendor: cost above the optimum (\%, median over records) under the level (L), tail (T), support (S), and coincidence (C) shifts. \textbf{(a)} Out-of-distribution cost at each method's selection rule. \textbf{(b)} In-distribution cost needed to halve SAA's out-of-distribution cost; $\cdot$: never reached; $^\dagger$: lowest in under 80\% of bootstrap resamples. Bold: lowest among the rows shown.}\label{tab:nv-main}
{\setlength{\tabcolsep}{2.6pt}
\begin{tabular}{lrrrr@{\hspace{5pt}}rrrr@{\hspace{5pt}}rrrr@{\hspace{5pt}}rrrr@{\hspace{5pt}}}
\toprule
 & \multicolumn{8}{c}{(a) OOD cost at SAA's ID budget} & \multicolumn{8}{c}{(b) ID cost to halve SAA's OOD cost} \\
\cmidrule(lr){2-9} \cmidrule(lr){10-17}
 & \multicolumn{4}{c}{$N = 16$} & \multicolumn{4}{c}{$N = 128$} & \multicolumn{4}{c}{$N = 16$} & \multicolumn{4}{c}{$N = 128$} \\
\cmidrule(lr){2-5} \cmidrule(lr){6-9} \cmidrule(lr){10-13} \cmidrule(lr){14-17}
Method & L & T & S & C & L & T & S & C & L & T & S & C & L & T & S & C \\
\midrule
SAA & 643 & 154 & 434 & 244 & 129 & 88 & 98 & 39 & $\cdot$ & $\cdot$ & $\cdot$ & $\cdot$ & $\cdot$ & $\cdot$ & $\cdot$ & $\cdot$ \\
KL-Rec & 647 & 155 & 435 & 227 & 141 & 89 & 111 & 39 & $\cdot$ & $\cdot$ & $\cdot$ & $\cdot$ & $\cdot$ & $\cdot$ & $\cdot$ & $\cdot$ \\
W-DRO & 80 & \textbf{39} & 92 & 154 & \textbf{24} & \textbf{45} & \textbf{30} & 18 & 55 & 68 & 55 & 97 & 12 & 42 & 12 & 20 \\
KL-OT & 88 & 43 & 99 & 159 & 45 & 57 & 44 & 25 & 57 & 64 & 57 & 97 & 12 & \textbf{29} & 12 & $\cdot$ \\
\midrule
\multicolumn{17}{l}{KL-Gen-P} \\
\quad Log-normal & 52 & 52 & 47 & \textbf{31} & 35 & 56 & 35 & \textbf{15} & 21 & 23 & 21 & 24 & \textbf{9}$^\dagger$ & $\cdot$ & \textbf{9}$^\dagger$ & \textbf{15} \\
\quad Flow & \textbf{52} & 52 & \textbf{43} & 33 & 41 & 58 & 35 & 18 & \textbf{19} & \textbf{19} & \textbf{19} & \textbf{23}$^\dagger$ & 11 & $\cdot$ & 9 & 19 \\
\quad Gaussian & 229 & 104 & 158 & 49 & 250 & 107 & 174 & 51 & 40 & $\cdot$ & 40 & 42 & $\cdot$ & $\cdot$ & $\cdot$ & $\cdot$ \\
KL-Gen (flow) & 229 & 103 & 152 & 49 & 206 & 101 & 138 & 46 & 36 & $\cdot$ & 36 & 33 & $\cdot$ & $\cdot$ & $\cdot$ & $\cdot$ \\
\bottomrule
\end{tabular}}

\end{table}

\paragraph{Record size and shift.}\looseness=-1  The optimal order depends on each product's top 1\% of demand, which lies above the record's maximum in 85\% of records at $N = 16$. Reweighting the record cannot reach this region, so KL-Rec costs as much as SAA at every $N$. The methods that extrapolate beyond the record can reach this region. At $N = 16$, where cross-validation keeps the flow at its log-normal initialization (App.~\ref{app:nv_nominals}), KL-Gen-P with the log-normal or the flow does so most cheaply, halving SAA's OOD cost at about a third of the ID cost that W-DRO and KL-OT need (Table~\ref{tab:nv-main}b). Which method leads at a fixed ID budget, however, depends on the shift (Table~\ref{tab:nv-main}a). W-DRO leads or ties under the tail shift at every $N$ and overtakes KL-Gen-P under the level and support shifts as $N$ grows. KL-Gen-P stays lowest under the coincidence shift because a fitted nominal places mass on joint shocks, which transport reaches only by moving every product of a record point at once.

\paragraph{The proposal and the nominal.}\looseness=-1  At the same 356 cost evaluations, KL-Gen-P cuts the out-of-distribution cost of KL-Gen by a factor of two to six under the level and support shifts for every nominal fitted to log demand other than the copula (Tables~\ref{tab:nv-full-16}--\ref{tab:nv-full-128}). This is because only about four of 356 draws from a nominal exceed a product's 99th percentile, whereas the proposal draws there directly (Sec.~\ref{sec:etas}). The copula is the exception, because its draws stay within the record's range whatever the latent proposal, and for it the proposal gains at most a factor of two. Nor does a larger generated pool substitute for sampling toward the worst case, since CSSC, which reduces a pool of 1{,}024 generated draws at 95 times the evaluations, costs as much as KL-Gen. Finally, the nominal's tail matters more than its exact family. With the proposal, the log-normal costs less out of distribution than the Gaussian in every cell, and less than the log-normal mixture, the family of the base distribution, under the level and support shifts up to $N = 64$ and under the coincidence shift at every $N$. These results support training generative models to reproduce the extremes of the process accurately \citep{jaini2020tails, laszkiewicz2022marginal, pandey2025heavy}.

\subsection{Capacity Expansion Planning} \label{sec:cep}

\paragraph{Problem.}\looseness=-1  A planner sizes wind, solar, gas, and battery capacity in 20 zones of the Texas system, and transmission between them, within a capital budget, then dispatches hourly, with a linearized unit commitment of gas plants, over a day of uncertain zonal demand and renewable availability ($D = 1{,}440$). Unserved load costs \$10{,}000/MWh, so cost is dominated by the few days on which high demand coincides with low renewable output across zones, and a record of a year or less rarely holds the worst of them.

\paragraph{Setup.}\looseness=-1  A record covers $k$ years of 2001--2010: every second day of one year ($k = \tfrac12$), a full year ($k = 1$), or, in App.~\ref{app:cep_results}, five years ($k = 5$). The learned nominal is a seasonally conditioned normalizing flow, fitted to the record alone and decoded through the record's marginal distributions (App.~\ref{app:generator}). Every method minimizes annualized capital cost plus the mean--CVaR of operating cost under the same capital budget ($B = 21.5$~bn\$/yr), holds at most $K = 48$ scenarios, and, if adaptive, re-targets every round. For the transport methods, KL-OT and W-DRO, the tables report the transport budget $\delta = 1$ (App.~\ref{app:cep_method}). Each plan is scored by the same objective, in bn\$/yr, on every day of 2011--2014 (ID), the four years after the record decade, and of 2020 (OOD), the most extreme year of 2001--2020 in the tail of demand (App.~\ref{app:cep_data}).

\begin{figure}[t]
    \centering
    \includegraphics[width=0.96\linewidth]{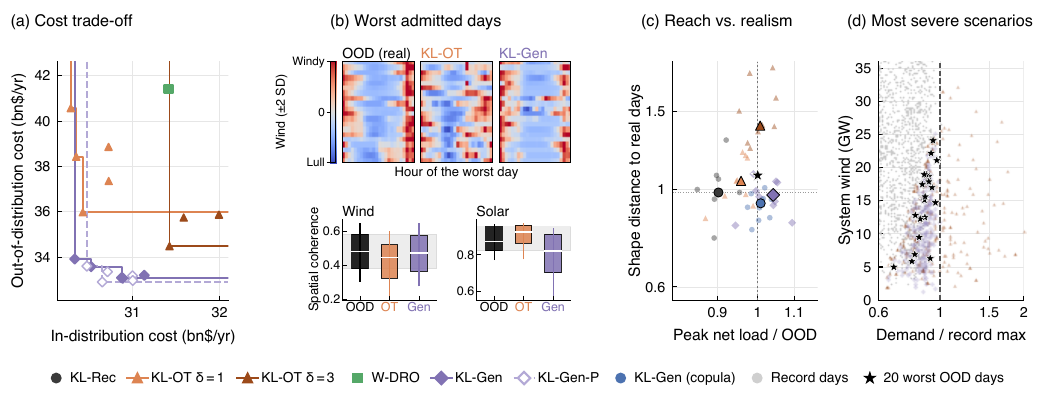}
    \caption{Capacity expansion at $k = \tfrac12$ (panels defined in App.~\ref{app:cep_eval}). \textbf{(a)} OOD against ID cost (bn\$/yr) of each setting, with Pareto frontiers; KL-Rec, CSSC, and SAA lie outside the range. \textbf{(b)} Zonal wind on the most severe OOD day and admitted scenarios (top), and their spatial coherence (bottom). \textbf{(c)} Severity against demand-rescaled distance to the nearest real day, for the most severe admitted scenarios. \textbf{(d)} Demand relative to its 2001--2010 maximum against system wind; stars: the 20 most severe OOD days.}
    \label{fig:cep}
\end{figure}

\paragraph{Frontiers across record lengths.}\looseness=-1  Adaptive sampling alone cuts cost well below that of a static set of record days. At $\rho = 0$, KL-Rec is far cheaper than SAA at every $k$ (Table~\ref{tab:cep-pareto}). At $k = \tfrac12$ and the five radius pairs of Table~\ref{tab:cep-pareto}, every setting of KL-Rec, and every setting of KL-OT except its cheapest in distribution, is dominated in median cost by a setting of KL-Gen (Figure~\ref{fig:cep}(a)). Over the three $k$, KL-Gen is cheaper than KL-OT out of distribution at all 15 radius pairs, at a small in-distribution premium for $k \ge 1$ (Table~\ref{tab:cep-paired}). The advantage narrows as the record grows: KL-Gen's out-of-distribution margin over KL-OT falls from 2.3--6.5~bn\$/yr at $k = \tfrac12$ to 0.07--0.76 at $k = 5$, and KL-Rec joins the Pareto front at $k = 5$. KL-OT reaches the front at every $k$ only through its cheapest in-distribution setting, and SAA, CSSC, and W-DRO never reach it. KL-Gen-P and KL-Gen with a seasonal Gaussian copula stay close to KL-Gen. The copula appears to capture what these decisions need, although the flow fits held-out days better (App.~\ref{app:generator}), and at this dimension the fitted proposal adds little over resampling (Sec.~\ref{sec:etas}).

\paragraph{What the generator extrapolates.}\looseness=-1  The most severe days of 2020 lie within the demand range of 2001--2010 and are severe because high demand coincides with low renewable output across zones (panel d), a pattern present even in a $k = \tfrac12$ record. The flow reproduces it: its most severe scenarios reach the OOD days' severity with spatial shapes as close to real days as the record's own severe days (panel c) and wind as spatially coherent as on real days (panel b).

\paragraph{What transport extrapolates.}\looseness=-1  KL-OT makes scenarios costly by inflating demand zone by zone, up to 2.4 times the highest system demand of 2001--2010 at $\delta = 3$ (panel d), whereas on real days each zone's demand moves with the system total. At $\delta = 1$ its most severe scenarios fall just short of the OOD days' severity, and at $\delta = 3$ they reach it only with shapes farther from real days than the record's own severe days (panel c). Widening the transport budget brings its OOD cost toward that of KL-Gen only at a higher ID cost (panel a). This premium is paid every year to hedge against days unlike those in the data.

\FloatBarrier
\section{Conclusion}
\looseness=-1 For two-stage stochastic programs with severe tail risk, we proposed centering a KL ball at a generative model of a short record, so that the adversary judges unobserved realizations by learned dependence rather than distance from the record, and solving the resulting problem on scenarios drawn adaptively toward its worst case. Across both case studies, we found that the learned nominal helps most when the record is short and the model class extrapolates along structure the record shows, such as the co-movement of demand across products or of renewable shortfalls across zones. The advantage fades as the record grows, and transport around the record hedges more cheaply against shifts the model class does not anticipate. As for the sampler, a fitted proposal adds most when the realizations driving the worst case are rare among the model's draws.

\paragraph{Limitations.}\looseness=-1 Both case studies have linear recourse, so integer and nonlinear convex recourse remain untested. On the theoretical side, the ball around $\mathbb Q_\theta$ contains a distribution $\mathbb P$, shifted or not, only if $\rho \ge \mathrm{KL}(\mathbb P \,\|\, \mathbb Q_\theta)$. This divergence is set by the model's misspecification, which the record does not reveal, and is infinite if the model assigns no mass where $\mathbb P$ has some, as nominals decoded through the record's empirical marginals do beyond its range. We therefore choose the radius empirically, without a finite-sample guarantee. Statistically, the master's objective is a biased estimate of its counterpart under $\mathbb Q_\theta$, and we do not show that the returned plan converges: under the alternation, it is a best response whose optimality gap on the held scenarios is bounded only after the fact (App.~\ref{app:optimization}).

\section*{AI Use Statement}
The authors received assistance from Claude Code in implementing the methods and baselines, producing the tables and figures from the data of the computational experiments, reviewing the literature, and editing and revising the text, and take full responsibility for the content of the paper.

\bibliographystyle{iclr2027_conference}
\bibliography{iclr2027_conference}

\appendix
\section{Newsvendor} \label{app:newsvendor}

\subsection{Problem} \label{app:nv_problem}

A retailer chooses order quantities $x \in \mathbb R^5_+$ at unit cost, $c(x) = \mathbf 1^\top x$, before demand $\xi \in \mathbb R^5_+$ is observed. Stock of product $i$ ships $y_{ij}$ units to demand for product $j$ along the arcs $\mathcal E = \{(j, j)\}_{j=1}^{5} \cup \{(j+1, j)\}_{j=1}^{4}$, and the recourse cost is
\begin{align} \label{eq:nv_recourse}
Q(x, \xi) = \min_{y, u, v \geq 0} \; & 0.5 \sum_{j=1}^{4} y_{j+1,j} + 100 \sum_{j=1}^{5} u_j - 0.25 \sum_{i=1}^{5} v_i \\
\text{s.t.} \; & \sum_{j : (i,j) \in \mathcal E} y_{ij} + v_i = x_i, \qquad \sum_{i : (i,j) \in \mathcal E} y_{ij} + u_j = \xi_j, \qquad i, j = 1, \dots, 5, \nonumber
\end{align}
where $u_j$ is unmet demand and $v_i$ leftover stock. The critical ratio is $(100 - 1)/(100 - 0.25) = 0.9925$, and recourse is relatively complete, since $y = 0$ is always feasible. The objective is $c(x)$ plus the mean--CVaR \eqref{eq:mean_cvar} of $Q(x, \xi)$ with $\beta = 0.25$ and $\alpha = 0.9$.

\subsection{Distributions and Records} \label{app:nv_laws}

\begin{figure}[htbp]
    \centering
    \includegraphics[width=\linewidth]{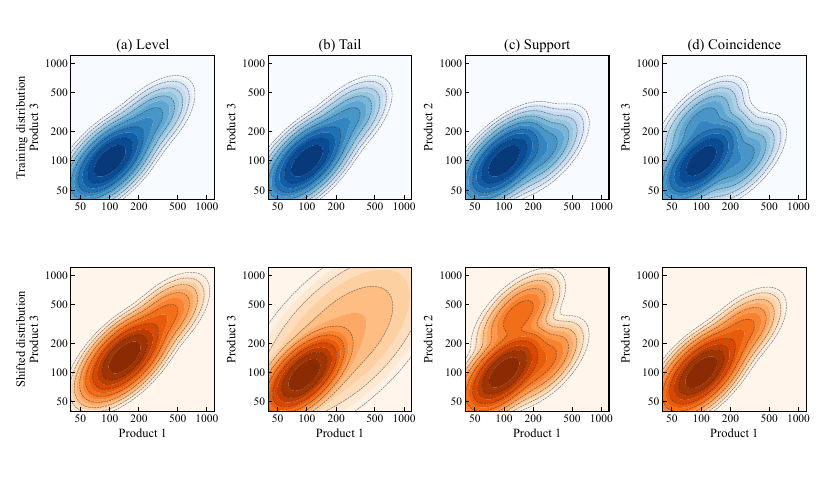}
    \caption{The four newsvendor shift families: joint densities of log demand for products 1 and 3 (1 and 2 in (c)) under the records' distribution (top) and the shifted distribution (bottom), with contours spanning 3.5 orders of magnitude so that the shocks are visible. \textbf{(a)} Level: the density translates. \textbf{(b)} Tail: the shock spreads into a wide, faint band. \textbf{(c)} Support: a new component appears where the records' distribution has almost no mass. \textbf{(d)} Coincidence: shocks to one product at a time become joint shocks, with unchanged marginals.}
    \label{fig:nv_shifts}
\end{figure}

Under the base distribution $\mathbb P_0$, log demand is $0.95\,\mathcal N(\mu, \Sigma) + 0.05\,\mathcal N(\mu + \Delta, \Sigma)$, with $\mu_j = \log 100$, standard deviations $0.35$, and correlations $0.6$. The shock $\Delta$ is $0.9$ on products 1, 3, and 5 and $0.225$ on products 2 and 4. Each shift family pairs the distribution the records are drawn from with a shifted distribution (Figure~\ref{fig:nv_shifts}).
\begin{description}[leftmargin=1em, itemsep=1pt, topsep=2pt]
    \item[Level.] Records from $\mathbb P_0$; the shifted distribution adds $0.35$ to every component of $\mu$.
    \item[Tail.] Records from $\mathbb P_0$; the shifted distribution multiplies the covariance of the shock component by 8, keeping expected demand unchanged.
    \item[Support.] Records from $\mathbb P_0$; the shifted distribution adds a third component of weight $0.05$ whose shock is $1.35$ on products 2 and 4 and $0.34$ on the others.
    \item[Coincidence.] Records from a mixture of a calm component of weight $0.75$ and five components of weight $0.05$, the $j$th adding $\Delta_j$ to product $j$ only; the shifted distribution is $\mathbb P_0$, with the same marginals.
\end{description}
There are 300 records at every $N$, nested across $N$, and the level, tail, and support families share them.

\subsection{Learned Nominals} \label{app:nv_nominals}

\begin{figure}[htbp]
    \centering
    \includegraphics{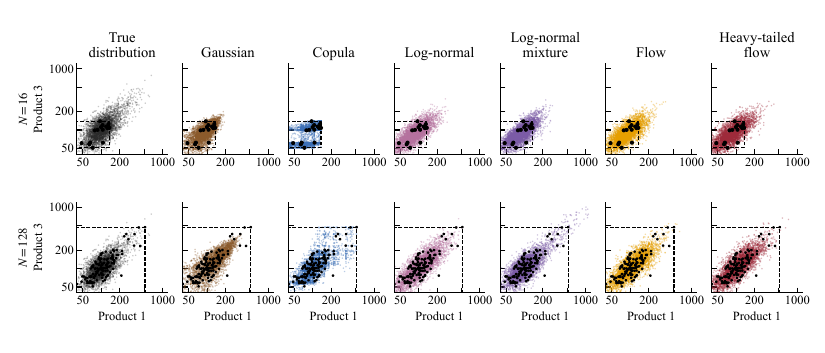}
    \caption{What each nominal can represent: 4{,}000 draws from each nominal fitted to one record of $N = 16$ (top) and one of $N = 128$ (bottom), shown for products 1 and 3 (black points: the record; dashed box: its range). The copula's draws stay within the record's range, while the other nominals extrapolate beyond it. Here, cross-validation keeps the flow at its log-normal initialization, and the heavy-tailed flow selects a log-$t$ with 30 degrees of freedom.}
    \label{fig:nv_approx}
\end{figure}

Each nominal, and a heavy-tailed variant of the flow, is fitted to all five products of a record and sampled as a map of $z \sim \mathcal N(0, I_5)$, so the weights $\varphi/r_t$ are exact whether or not the map is invertible (Sec.~\ref{sec:tilt}). Figure~\ref{fig:nv_approx} compares them on single records.
\begin{description}[leftmargin=1em, itemsep=1pt, topsep=2pt]
    \item[Gaussian.] Sample mean and covariance of raw demand, with draws clipped at zero.
    \item[Copula.] The correlation of the normal scores $\Phi^{-1}((m + \tfrac12)/N)$ of each coordinate's rank $m \in \{0, \dots, N-1\}$ in the record, where $\Phi$ is the standard normal distribution function. Draws are decoded through the record's empirical quantiles, so they stay within its range.
    \item[Log-normal.] A maximum-likelihood Gaussian on log demand.
    \item[Log-normal mixture.] Two components fitted by expectation maximization, the family of the base distribution, sampled through the inverse of its Rosenblatt transform.
    \item[Flow.] A RealNVP flow \citep{dinh_density_2017} on whitened log demand, with six affine coupling layers of width 16, initialized at the log-normal and fitted by maximum likelihood with Adam. Its number of steps (0 to 400) and weight decay are chosen by five-fold cross-validation, which keeps zero steps on every record at $N = 16$. There the flow remains a log-normal, but with the unbiased covariance (divisor $N-1$) rather than the maximum-likelihood one (divisor $N$), so its rows differ from the log-normal's by this 16/15 wider covariance and by independent sampler draws.
    \item[Heavy-tailed flow.] A tail-adaptive flow after \citet{jaini2020tails}: the same flow on a Student-$t$ base with a learned number of degrees of freedom per coordinate, obtained by mapping each latent coordinate through $\Phi$ and then the Student-$t$ quantile function. Cross-validation chooses among the log-normal, a log-$t$, and the $t$-base flow.
\end{description}

\subsection{Methods} \label{app:nv_methods}

Table~\ref{tab:nv_methods} summarizes the methods. Methods with a KL ball are run at every radius of the grid $\rho \in \{0, 0.003, 0.005, 0.007, 0.01, 0.015, 0.02, 0.03, 0.05, 0.1, 0.2, 0.4, 0.8\}$, and methods that depend on a decision use the risk-neutral decision $\hat x$ on the record and tilt at $\rho_{\mathrm{ref}} = 0.8$.
\begin{description}[leftmargin=1em, itemsep=1pt, topsep=2pt]
    \item[KL-Rec and SAA.] KL-DRO on the empirical distribution \citep{hu2013kullback}. SAA is KL-Rec at $\rho = 0$.
    \item[KL-Gen-P and the oracle.] One round of the sampler at $\hat x$: $M = 256$ latent draws are evaluated, a Gaussian is fitted to their tilt, with its covariance shrunk toward its diagonal with weight 0.25, and $n = 100$ scenarios are drawn from the defensive mixture with $a = 0.9$, once per record for all radii. The oracle applies the same sampler to the true distribution and serves only as a reference.
    \item[KL-Gen.] Resamples 100 of 356 draws of the nominal by the tilt, with replacement, matching the $256 + 100$ recourse evaluations of the sampler.
    \item[CSSC.] Reduces 1{,}024 draws of the log-normal or the log-normal mixture to $K = 100$ medoids weighted by cluster mass, clustering each draw's costs under $\hat x$ and 32 single-scenario decisions with $k$-medoids, a scalable variant of the program of \citet{keutchayan_problem-driven_2023}.
    \item[Transport step.] W-DRO and KL-OT move each record point $\xi^0$ to an approximate maximizer of $Q(x, \xi) - \gamma \|\omega(\xi) - \omega(\xi^0)\|_2^2$, the Lagrangian of a type-2 Wasserstein ball, with $\omega$ the standardized log demand, by damped fixed-point steps on the recourse duals. The multiplier $\gamma$ is set at $\hat x$ so that the mean squared displacement is $\varepsilon$ times the median squared nearest-neighbor spacing of the record, and we call $\varepsilon$ the displacement. Both methods alternate three rounds of moving the points and solving the master.
    \item[W-DRO.] Moves all $N$ record points, with $\varepsilon \in \{0.01, 0.03, 0.1, 0.3, 1, 3, 10\}$, and has no KL ball.
    \item[KL-OT.] Moves $\min(100, N)$ record points selected by the tilt, at the displacements of W-DRO, within the ball at every radius of the grid.
\end{description}
\paragraph{Parity.} All methods share the records, the evaluation distributions, the selection rule, and the master. KL-Gen-P and KL-Gen evaluate the recourse 356 times per record, and KL-Gen's master holds a median of 69 distinct scenarios against the 100 of KL-Gen-P. CSSC evaluates it 33 times per pool draw, and W-DRO and KL-OT about 1{,}300 times per setting at $N = 16$ and 8{,}400 to 9{,}700 at $N = 128$. KL-OT also chooses among 91 settings, against 13 for the other KL methods and 7 for W-DRO. The differences in evaluations and in KL-OT's grid favor the baselines.

\begin{table}[htbp]
\centering
\caption{Newsvendor methods. KL-Gen-P and KL-Gen are run with each nominal of Figure~\ref{fig:nv_approx}, and CSSC with the log-normal and the log-normal mixture. Weights are normalized to sum to one.}
\label{tab:nv_methods}
\small
\begin{tabular}{lp{0.42\linewidth}ll}
\toprule
Method & Scenarios in the master & Weights & Radius \\
\midrule
SAA & the $N$ record points & uniform & $0$ \\
KL-Rec & the $N$ record points & uniform & grid \\
KL-Gen-P, oracle & $n = 100$ draws of the sampler & $\varphi / r$ & grid \\
KL-Gen & 356 draws of the nominal, resampled 100 times by the tilt & $c_i/(356\,\tau_i)$ & grid \\
CSSC & $K = 100$ medoids of 1{,}024 draws of the nominal & cluster mass & grid \\
W-DRO & the $N$ record points, moved & uniform & $0$ \\
KL-OT & $\min(100, N)$ record points selected by the tilt, moved & uniform & grid \\
\bottomrule
\end{tabular}
\end{table}

\subsection{Solving the Master} \label{app:nv_master}

The master is a linear program at $\rho = 0$, solved by Gurobi \citep{gurobi}, and otherwise an exponential-cone program solved by Clarabel \citep{goulart2024clarabel}. A Clarabel solution is accepted if its value agrees with the objective recomputed at its decision to $10^{-5}$ relative and is no worse than the solution for $\rho = 0$. Otherwise, a sequential least squares programming solve starts from the solution for $\rho = 0$. This fallback served 10\% of the master solves with $\rho > 0$, and 22\% of those of KL-Gen-P.

\subsection{Selection and Evaluation} \label{app:nv_eval}

\paragraph{Selection rule.} Records are split into odd- and even-numbered halves. On each half, each method's setting is chosen as the largest whose mean in-distribution cost is within 5 percentage points of that of SAA, or the cheapest if none is, and it is scored on the other half, so the shifted distribution is never consulted. Settings are ordered by radius, by displacement for W-DRO, and by displacement and then radius for KL-OT. Table~\ref{tab:nv-main}(b) instead reads the frontiers of Figure~\ref{fig:nv-frontiers} at a fixed target: the lowest median in-distribution cost among a method's settings whose median out-of-distribution cost is at most half that of SAA, with the stability of the lowest method assessed over 200 bootstrap resamples of the records. Under relative rules that admit in-distribution costs up to 1.10 or 1.25 times that of SAA, the lowest method is unchanged in 7 of the 8 columns of Table~\ref{tab:nv-main}(a) at 1.10, where the log-normal replaces W-DRO at $N = 128$ under the tail shift, and in all 8 at 1.25.

\paragraph{Scoring.} In- and out-of-distribution costs are the objective of App.~\ref{app:nv_problem} under the records' distribution and the shifted distribution, as a percentage above the optimum under that distribution, computed on a scrambled Sobol quadrature with 4{,}096 points per mixture component. Tables report the median over records with the standard error $1.2533\,\hat\sigma/\sqrt n$, where $\hat\sigma$ is the interquartile range divided by $1.349$.

\subsection{Full Results} \label{app:nv_results}

\begin{figure}[htbp]
    \centering
    \includegraphics[width=\linewidth]{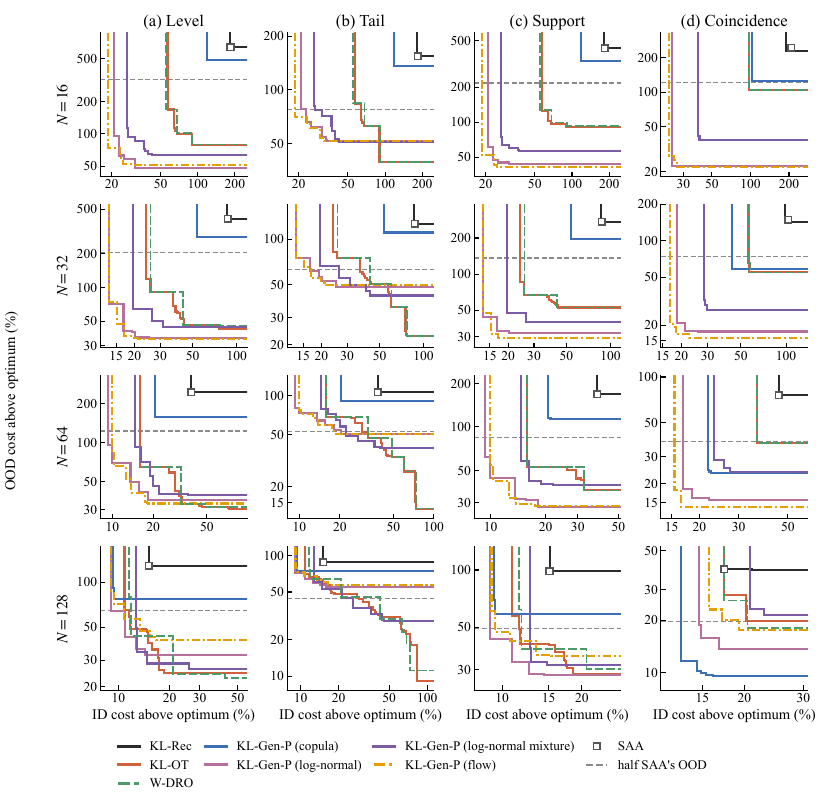}
    \caption{Newsvendor frontiers: out-of-distribution against in-distribution cost above the optimum (\%, median over records) as each method's settings vary. Each line is a method's Pareto frontier (lower left is better). Rows are record sizes and columns shifts. The open square is SAA, and the dashed line is half of SAA's out-of-distribution cost, the target of Table~\ref{tab:nv-main}(b).}
    \label{fig:nv-frontiers}
\end{figure}

Figure~\ref{fig:nv-frontiers} and Tables~\ref{tab:nv-full-16}--\ref{tab:nv-full-128} give the full results.

\begin{table}[htbp]\centering\scriptsize
\caption{Newsvendor, every method, $N = 16$ (300 records): cost above the optimum (\%) under the level (L), tail (T), support (S), and coincidence (C) shifts. (a) Median out-of-distribution cost ($\pm$ 2 SE) at each method's selection rule. ID cost at $\rho = 0$: median in-distribution cost of the unhedged fit (none for W-DRO and KL-OT). (b) Lowest median in-distribution cost at which a setting halves SAA's median out-of-distribution cost; $\cdot$: never. Bold: lowest per column, excluding the oracle; $^\dagger$: lowest in fewer than 80\% of bootstrap resamples.}\label{tab:nv-full-16}

{\setlength{\tabcolsep}{3.4pt}
\begin{tabular}{lrrrr@{\hspace{7pt}}r@{\hspace{7pt}}rrrr}
\toprule
 & \multicolumn{4}{c}{(a) OOD cost at SAA's ID budget} & ID cost & \multicolumn{4}{c}{(b) ID cost to halve SAA's OOD} \\
\cmidrule(lr){2-5} \cmidrule(lr){6-6} \cmidrule(lr){7-10}
Method & L & T & S & C & $\rho = 0$ & L & T & S & C \\
\midrule
SAA & 643\,{\tiny$\pm$59} & 154\,{\tiny$\pm$8} & 434\,{\tiny$\pm$39} & 244\,{\tiny$\pm$21} & 184 & $\cdot$ & $\cdot$ & $\cdot$ & $\cdot$ \\
KL-Rec & 647\,{\tiny$\pm$55} & 155\,{\tiny$\pm$7} & 435\,{\tiny$\pm$36} & 227\,{\tiny$\pm$20} & 184 & $\cdot$ & $\cdot$ & $\cdot$ & $\cdot$ \\
W-DRO & 80\,{\tiny$\pm$10} & 39\,{\tiny$\pm$4} & 92\,{\tiny$\pm$10} & 154\,{\tiny$\pm$20} & -- & 55 & 68 & 55 & 97 \\
KL-OT & 88\,{\tiny$\pm$11} & 43\,{\tiny$\pm$4} & 99\,{\tiny$\pm$11} & 159\,{\tiny$\pm$20} & -- & 57 & 64 & 57 & 97 \\
\multicolumn{10}{l}{\textit{Gaussian}} \\
\quad KL-Gen-P & 229\,{\tiny$\pm$31} & 104\,{\tiny$\pm$4} & 158\,{\tiny$\pm$19} & 49\,{\tiny$\pm$7} & 115 & 40 & $\cdot$ & 40 & 42 \\
\quad KL-Gen & 480\,{\tiny$\pm$43} & 135\,{\tiny$\pm$5} & 316\,{\tiny$\pm$27} & 128\,{\tiny$\pm$12} & 116 & $\cdot$ & $\cdot$ & $\cdot$ & $\cdot$ \\
\multicolumn{10}{l}{\textit{Copula}} \\
\quad KL-Gen-P & 490\,{\tiny$\pm$46} & 136\,{\tiny$\pm$6} & 338\,{\tiny$\pm$30} & 126\,{\tiny$\pm$12} & 125 & $\cdot$ & $\cdot$ & $\cdot$ & $\cdot$ \\
\quad KL-Gen & 503\,{\tiny$\pm$48} & 137\,{\tiny$\pm$6} & 344\,{\tiny$\pm$30} & 133\,{\tiny$\pm$13} & 127 & $\cdot$ & $\cdot$ & $\cdot$ & $\cdot$ \\
\multicolumn{10}{l}{\textit{Log-normal}} \\
\quad KL-Gen-P & 52\,{\tiny$\pm$11} & 52\,{\tiny$\pm$6} & 47\,{\tiny$\pm$8} & \textbf{31}\,{\tiny$\pm$5} & 48 & 21 & 23 & 21 & 24 \\
\quad KL-Gen & 238\,{\tiny$\pm$37} & 105\,{\tiny$\pm$5} & 162\,{\tiny$\pm$23} & 56\,{\tiny$\pm$9} & 48 & 43 & $\cdot$ & 43 & 40 \\
\quad CSSC & 240\,{\tiny$\pm$35} & 104\,{\tiny$\pm$5} & 163\,{\tiny$\pm$22} & 66\,{\tiny$\pm$9} & 57 & 44 & $\cdot$ & 44 & 45 \\
\multicolumn{10}{l}{\textit{Log-normal mixture}} \\
\quad KL-Gen-P & 65\,{\tiny$\pm$13} & 52\,{\tiny$\pm$6} & 59\,{\tiny$\pm$10} & 44\,{\tiny$\pm$8} & 41 & 26 & 27 & 26 & 39 \\
\quad KL-Gen & 216\,{\tiny$\pm$34} & 102\,{\tiny$\pm$5} & 151\,{\tiny$\pm$20} & 59\,{\tiny$\pm$8} & 41 & 40 & $\cdot$ & 40 & 49 \\
\quad CSSC & 193\,{\tiny$\pm$35} & 97\,{\tiny$\pm$6} & 136\,{\tiny$\pm$24} & 63\,{\tiny$\pm$9} & 49 & 39 & $\cdot$ & 39 & 52 \\
\multicolumn{10}{l}{\textit{Flow}} \\
\quad KL-Gen-P & \textbf{52}\,{\tiny$\pm$9} & 52\,{\tiny$\pm$5} & \textbf{43}\,{\tiny$\pm$7} & 33\,{\tiny$\pm$6} & 41 & \textbf{19} & \textbf{19} & \textbf{19} & \textbf{23}$^\dagger$ \\
\quad KL-Gen & 229\,{\tiny$\pm$34} & 103\,{\tiny$\pm$5} & 152\,{\tiny$\pm$21} & 49\,{\tiny$\pm$7} & 38 & 36 & $\cdot$ & 36 & 33 \\
\multicolumn{10}{l}{\textit{Heavy-tailed flow}} \\
\quad KL-Gen-P & 55\,{\tiny$\pm$11} & \textbf{39}\,{\tiny$\pm$4} & 54\,{\tiny$\pm$11} & 49\,{\tiny$\pm$8} & 28 & 22 & 22 & 22 & 25 \\
\quad KL-Gen & 178\,{\tiny$\pm$29} & 96\,{\tiny$\pm$5} & 126\,{\tiny$\pm$20} & 39\,{\tiny$\pm$6} & 36 & 31 & $\cdot$ & 31 & 31 \\
\midrule
\textit{Oracle (reference)} & 18\,{\tiny$\pm$3} & 21\,{\tiny$\pm$2} & 32\,{\tiny$\pm$3} & 22\,{\tiny$\pm$3} & 3 &  &  &  &  \\
\bottomrule
\end{tabular}}

\end{table}

\begin{table}[htbp]\centering\scriptsize
\caption{Newsvendor, every method, $N = 32$ (300 records); as Table~\ref{tab:nv-full-16}.}\label{tab:nv-full-32}

{\setlength{\tabcolsep}{3.4pt}
\begin{tabular}{lrrrr@{\hspace{7pt}}r@{\hspace{7pt}}rrrr}
\toprule
 & \multicolumn{4}{c}{(a) OOD cost at SAA's ID budget} & ID cost & \multicolumn{4}{c}{(b) ID cost to halve SAA's OOD} \\
\cmidrule(lr){2-5} \cmidrule(lr){6-6} \cmidrule(lr){7-10}
Method & L & T & S & C & $\rho = 0$ & L & T & S & C \\
\midrule
SAA & 412\,{\tiny$\pm$41} & 126\,{\tiny$\pm$5} & 273\,{\tiny$\pm$26} & 149\,{\tiny$\pm$14} & 87 & $\cdot$ & $\cdot$ & $\cdot$ & $\cdot$ \\
KL-Rec & 412\,{\tiny$\pm$41} & 126\,{\tiny$\pm$5} & 284\,{\tiny$\pm$27} & 143\,{\tiny$\pm$13} & 87 & $\cdot$ & $\cdot$ & $\cdot$ & $\cdot$ \\
W-DRO & 46\,{\tiny$\pm$5} & \textbf{23}\,{\tiny$\pm$3} & 57\,{\tiny$\pm$6} & 85\,{\tiny$\pm$11} & -- & 26 & 43 & 26 & 56 \\
KL-OT & 46\,{\tiny$\pm$5} & 24\,{\tiny$\pm$3} & 61\,{\tiny$\pm$6} & 98\,{\tiny$\pm$12} & -- & 24 & 37 & 24 & 56 \\
\multicolumn{10}{l}{\textit{Gaussian}} \\
\quad KL-Gen-P & 222\,{\tiny$\pm$27} & 104\,{\tiny$\pm$4} & 159\,{\tiny$\pm$17} & 50\,{\tiny$\pm$7} & 93 & $\cdot$ & $\cdot$ & $\cdot$ & 37 \\
\quad KL-Gen & 414\,{\tiny$\pm$30} & 127\,{\tiny$\pm$3} & 278\,{\tiny$\pm$19} & 119\,{\tiny$\pm$9} & 90 & $\cdot$ & $\cdot$ & $\cdot$ & $\cdot$ \\
\multicolumn{10}{l}{\textit{Copula}} \\
\quad KL-Gen-P & 285\,{\tiny$\pm$29} & 111\,{\tiny$\pm$4} & 195\,{\tiny$\pm$18} & 58\,{\tiny$\pm$6} & 61 & $\cdot$ & $\cdot$ & $\cdot$ & 44 \\
\quad KL-Gen & 313\,{\tiny$\pm$29} & 114\,{\tiny$\pm$4} & 214\,{\tiny$\pm$19} & 66\,{\tiny$\pm$8} & 60 & $\cdot$ & $\cdot$ & $\cdot$ & 53 \\
\multicolumn{10}{l}{\textit{Log-normal}} \\
\quad KL-Gen-P & \textbf{36}\,{\tiny$\pm$7} & 49\,{\tiny$\pm$5} & 35\,{\tiny$\pm$5} & \textbf{23}\,{\tiny$\pm$3} & 37 & \textbf{13}$^\dagger$ & 17 & \textbf{13}$^\dagger$ & 19 \\
\quad KL-Gen & 190\,{\tiny$\pm$24} & 98\,{\tiny$\pm$4} & 138\,{\tiny$\pm$16} & 44\,{\tiny$\pm$7} & 33 & 28 & $\cdot$ & 28 & 29 \\
\quad CSSC & 223\,{\tiny$\pm$26} & 103\,{\tiny$\pm$4} & 151\,{\tiny$\pm$18} & 51\,{\tiny$\pm$7} & 44 & $\cdot$ & $\cdot$ & $\cdot$ & 35 \\
\multicolumn{10}{l}{\textit{Log-normal mixture}} \\
\quad KL-Gen-P & 49\,{\tiny$\pm$10} & 43\,{\tiny$\pm$5} & 51\,{\tiny$\pm$9} & 39\,{\tiny$\pm$5} & 30 & 20 & 26 & 20 & 28 \\
\quad KL-Gen & 168\,{\tiny$\pm$26} & 94\,{\tiny$\pm$5} & 124\,{\tiny$\pm$17} & 48\,{\tiny$\pm$7} & 32 & 28 & $\cdot$ & 28 & 35 \\
\quad CSSC & 160\,{\tiny$\pm$28} & 90\,{\tiny$\pm$5} & 119\,{\tiny$\pm$18} & 54\,{\tiny$\pm$6} & 32 & 30 & $\cdot$ & 30 & 38 \\
\multicolumn{10}{l}{\textit{Flow}} \\
\quad KL-Gen-P & 41\,{\tiny$\pm$9} & 55\,{\tiny$\pm$5} & \textbf{33}\,{\tiny$\pm$6} & 24\,{\tiny$\pm$3} & 34 & 13 & 17 & 13 & \textbf{17} \\
\quad KL-Gen & 195\,{\tiny$\pm$28} & 98\,{\tiny$\pm$4} & 135\,{\tiny$\pm$18} & 45\,{\tiny$\pm$7} & 30 & 27 & $\cdot$ & 27 & 29 \\
\multicolumn{10}{l}{\textit{Heavy-tailed flow}} \\
\quad KL-Gen-P & 44\,{\tiny$\pm$8} & 37\,{\tiny$\pm$5} & 46\,{\tiny$\pm$7} & 43\,{\tiny$\pm$6} & 24 & 14 & \textbf{17}$^\dagger$ & 14 & 20 \\
\quad KL-Gen & 170\,{\tiny$\pm$24} & 94\,{\tiny$\pm$4} & 120\,{\tiny$\pm$16} & 34\,{\tiny$\pm$5} & 28 & 25 & $\cdot$ & 25 & 25 \\
\midrule
\textit{Oracle (reference)} & 19\,{\tiny$\pm$3} & 21\,{\tiny$\pm$2} & 32\,{\tiny$\pm$3} & 23\,{\tiny$\pm$3} & 3 &  &  &  &  \\
\bottomrule
\end{tabular}}

\end{table}

\begin{table}[htbp]\centering\scriptsize
\caption{Newsvendor, every method, $N = 64$ (300 records); as Table~\ref{tab:nv-full-16}.}\label{tab:nv-full-64}

{\setlength{\tabcolsep}{3.4pt}
\begin{tabular}{lrrrr@{\hspace{7pt}}r@{\hspace{7pt}}rrrr}
\toprule
 & \multicolumn{4}{c}{(a) OOD cost at SAA's ID budget} & ID cost & \multicolumn{4}{c}{(b) ID cost to halve SAA's OOD} \\
\cmidrule(lr){2-5} \cmidrule(lr){6-6} \cmidrule(lr){7-10}
Method & L & T & S & C & $\rho = 0$ & L & T & S & C \\
\midrule
SAA & 245\,{\tiny$\pm$30} & 107\,{\tiny$\pm$4} & 169\,{\tiny$\pm$18} & 76\,{\tiny$\pm$10} & 38 & $\cdot$ & $\cdot$ & $\cdot$ & $\cdot$ \\
KL-Rec & 264\,{\tiny$\pm$30} & 108\,{\tiny$\pm$4} & 182\,{\tiny$\pm$18} & 78\,{\tiny$\pm$8} & 38 & $\cdot$ & $\cdot$ & $\cdot$ & $\cdot$ \\
W-DRO & \textbf{31}\,{\tiny$\pm$4} & \textbf{34}\,{\tiny$\pm$3} & 44\,{\tiny$\pm$5} & 37\,{\tiny$\pm$4} & -- & 16 & 32 & 16 & 36 \\
KL-OT & 39\,{\tiny$\pm$8} & 45\,{\tiny$\pm$4} & 48\,{\tiny$\pm$6} & 51\,{\tiny$\pm$7} & -- & 16 & 29 & 16 & 36 \\
\multicolumn{10}{l}{\textit{Gaussian}} \\
\quad KL-Gen-P & 245\,{\tiny$\pm$20} & 107\,{\tiny$\pm$3} & 172\,{\tiny$\pm$13} & 55\,{\tiny$\pm$6} & 90 & $\cdot$ & $\cdot$ & $\cdot$ & $\cdot$ \\
\quad KL-Gen & 406\,{\tiny$\pm$21} & 126\,{\tiny$\pm$2} & 267\,{\tiny$\pm$13} & 113\,{\tiny$\pm$8} & 87 & $\cdot$ & $\cdot$ & $\cdot$ & $\cdot$ \\
\multicolumn{10}{l}{\textit{Copula}} \\
\quad KL-Gen-P & 157\,{\tiny$\pm$18} & 92\,{\tiny$\pm$3} & 115\,{\tiny$\pm$11} & 24\,{\tiny$\pm$4} & 30 & $\cdot$ & $\cdot$ & $\cdot$ & 22 \\
\quad KL-Gen & 205\,{\tiny$\pm$21} & 101\,{\tiny$\pm$3} & 151\,{\tiny$\pm$14} & 39\,{\tiny$\pm$5} & 33 & $\cdot$ & $\cdot$ & $\cdot$ & 26 \\
\multicolumn{10}{l}{\textit{Log-normal}} \\
\quad KL-Gen-P & 39\,{\tiny$\pm$7} & 52\,{\tiny$\pm$4} & 31\,{\tiny$\pm$5} & \textbf{21}\,{\tiny$\pm$3} & 35 & \textbf{9}$^\dagger$ & 20 & \textbf{9}$^\dagger$ & 17 \\
\quad KL-Gen & 204\,{\tiny$\pm$21} & 101\,{\tiny$\pm$3} & 145\,{\tiny$\pm$13} & 43\,{\tiny$\pm$6} & 34 & $\cdot$ & $\cdot$ & $\cdot$ & $\cdot$ \\
\quad CSSC & 209\,{\tiny$\pm$21} & 101\,{\tiny$\pm$3} & 151\,{\tiny$\pm$14} & 45\,{\tiny$\pm$6} & 42 & $\cdot$ & $\cdot$ & $\cdot$ & $\cdot$ \\
\multicolumn{10}{l}{\textit{Log-normal mixture}} \\
\quad KL-Gen-P & 41\,{\tiny$\pm$9} & 45\,{\tiny$\pm$5} & 42\,{\tiny$\pm$6} & 35\,{\tiny$\pm$5} & 22 & 15 & 22 & 15 & 23 \\
\quad KL-Gen & 143\,{\tiny$\pm$21} & 89\,{\tiny$\pm$4} & 109\,{\tiny$\pm$13} & 42\,{\tiny$\pm$6} & 25 & $\cdot$ & $\cdot$ & $\cdot$ & $\cdot$ \\
\quad CSSC & 127\,{\tiny$\pm$21} & 85\,{\tiny$\pm$5} & 99\,{\tiny$\pm$13} & 44\,{\tiny$\pm$5} & 26 & 24 & $\cdot$ & $\cdot$ & $\cdot$ \\
\multicolumn{10}{l}{\textit{Flow}} \\
\quad KL-Gen-P & 34\,{\tiny$\pm$6} & 51\,{\tiny$\pm$4} & \textbf{31}\,{\tiny$\pm$4} & 22\,{\tiny$\pm$3} & 34 & 10 & \textbf{17} & 10 & \textbf{15} \\
\quad KL-Gen & 204\,{\tiny$\pm$22} & 100\,{\tiny$\pm$3} & 144\,{\tiny$\pm$14} & 47\,{\tiny$\pm$6} & 35 & $\cdot$ & $\cdot$ & $\cdot$ & $\cdot$ \\
\multicolumn{10}{l}{\textit{Heavy-tailed flow}} \\
\quad KL-Gen-P & 35\,{\tiny$\pm$7} & 43\,{\tiny$\pm$5} & 35\,{\tiny$\pm$5} & 28\,{\tiny$\pm$4} & 25 & 12 & 28 & 12 & 18 \\
\quad KL-Gen & 177\,{\tiny$\pm$20} & 96\,{\tiny$\pm$3} & 128\,{\tiny$\pm$13} & 35\,{\tiny$\pm$5} & 30 & $\cdot$ & $\cdot$ & $\cdot$ & 23 \\
\midrule
\textit{Oracle (reference)} & 19\,{\tiny$\pm$3} & 22\,{\tiny$\pm$2} & 30\,{\tiny$\pm$3} & 25\,{\tiny$\pm$3} & 3 &  &  &  &  \\
\bottomrule
\end{tabular}}

\end{table}

\begin{table}[htbp]\centering\scriptsize
\caption{Newsvendor, every method, $N = 128$ (300 records); as Table~\ref{tab:nv-full-16}.}\label{tab:nv-full-128}

{\setlength{\tabcolsep}{3.4pt}
\begin{tabular}{lrrrr@{\hspace{7pt}}r@{\hspace{7pt}}rrrr}
\toprule
 & \multicolumn{4}{c}{(a) OOD cost at SAA's ID budget} & ID cost & \multicolumn{4}{c}{(b) ID cost to halve SAA's OOD} \\
\cmidrule(lr){2-5} \cmidrule(lr){6-6} \cmidrule(lr){7-10}
Method & L & T & S & C & $\rho = 0$ & L & T & S & C \\
\midrule
SAA & 129\,{\tiny$\pm$17} & 88\,{\tiny$\pm$3} & 98\,{\tiny$\pm$11} & 39\,{\tiny$\pm$5} & 15 & $\cdot$ & $\cdot$ & $\cdot$ & $\cdot$ \\
KL-Rec & 141\,{\tiny$\pm$19} & 89\,{\tiny$\pm$3} & 111\,{\tiny$\pm$13} & 39\,{\tiny$\pm$4} & 15 & $\cdot$ & $\cdot$ & $\cdot$ & $\cdot$ \\
W-DRO & \textbf{24}\,{\tiny$\pm$5} & \textbf{45}\,{\tiny$\pm$3} & \textbf{30}\,{\tiny$\pm$3} & 18\,{\tiny$\pm$3} & -- & 12 & 42 & 12 & 20 \\
KL-OT & 45\,{\tiny$\pm$8} & 57\,{\tiny$\pm$4} & 44\,{\tiny$\pm$6} & 25\,{\tiny$\pm$3} & -- & 12 & 29 & 12 & $\cdot$ \\
\multicolumn{10}{l}{\textit{Gaussian}} \\
\quad KL-Gen-P & 250\,{\tiny$\pm$18} & 107\,{\tiny$\pm$2} & 174\,{\tiny$\pm$11} & 51\,{\tiny$\pm$4} & 81 & $\cdot$ & $\cdot$ & $\cdot$ & $\cdot$ \\
\quad KL-Gen & 389\,{\tiny$\pm$22} & 124\,{\tiny$\pm$2} & 261\,{\tiny$\pm$13} & 103\,{\tiny$\pm$6} & 82 & $\cdot$ & $\cdot$ & $\cdot$ & $\cdot$ \\
\multicolumn{10}{l}{\textit{Copula}} \\
\quad KL-Gen-P & 77\,{\tiny$\pm$9} & 75\,{\tiny$\pm$3} & 60\,{\tiny$\pm$7} & \textbf{10}\,{\tiny$\pm$1} & 15 & $\cdot$ & $\cdot$ & $\cdot$ & \textbf{13} \\
\quad KL-Gen & 151\,{\tiny$\pm$16} & 91\,{\tiny$\pm$3} & 108\,{\tiny$\pm$11} & 21\,{\tiny$\pm$3} & 18 & $\cdot$ & $\cdot$ & $\cdot$ & $\cdot$ \\
\multicolumn{10}{l}{\textit{Log-normal}} \\
\quad KL-Gen-P & 35\,{\tiny$\pm$7} & 56\,{\tiny$\pm$4} & 35\,{\tiny$\pm$5} & 15\,{\tiny$\pm$2} & 29 & \textbf{9}$^\dagger$ & $\cdot$ & \textbf{9}$^\dagger$ & 15 \\
\quad KL-Gen & 189\,{\tiny$\pm$17} & 99\,{\tiny$\pm$3} & 134\,{\tiny$\pm$12} & 40\,{\tiny$\pm$5} & 34 & $\cdot$ & $\cdot$ & $\cdot$ & $\cdot$ \\
\quad CSSC & 202\,{\tiny$\pm$18} & 101\,{\tiny$\pm$3} & 140\,{\tiny$\pm$12} & 41\,{\tiny$\pm$5} & 40 & $\cdot$ & $\cdot$ & $\cdot$ & $\cdot$ \\
\multicolumn{10}{l}{\textit{Log-normal mixture}} \\
\quad KL-Gen-P & 34\,{\tiny$\pm$8} & 54\,{\tiny$\pm$5} & 35\,{\tiny$\pm$5} & 23\,{\tiny$\pm$4} & 13 & 13 & \textbf{26} & 13 & $\cdot$ \\
\quad KL-Gen & 106\,{\tiny$\pm$17} & 82\,{\tiny$\pm$4} & 85\,{\tiny$\pm$12} & 39\,{\tiny$\pm$5} & 19 & $\cdot$ & $\cdot$ & $\cdot$ & $\cdot$ \\
\quad CSSC & 96\,{\tiny$\pm$16} & 79\,{\tiny$\pm$4} & 81\,{\tiny$\pm$11} & 36\,{\tiny$\pm$4} & 17 & $\cdot$ & $\cdot$ & $\cdot$ & $\cdot$ \\
\multicolumn{10}{l}{\textit{Flow}} \\
\quad KL-Gen-P & 41\,{\tiny$\pm$7} & 58\,{\tiny$\pm$4} & 35\,{\tiny$\pm$5} & 18\,{\tiny$\pm$3} & 33 & 11 & $\cdot$ & 9 & 19 \\
\quad KL-Gen & 206\,{\tiny$\pm$20} & 101\,{\tiny$\pm$3} & 138\,{\tiny$\pm$13} & 46\,{\tiny$\pm$5} & 31 & $\cdot$ & $\cdot$ & $\cdot$ & $\cdot$ \\
\multicolumn{10}{l}{\textit{Heavy-tailed flow}} \\
\quad KL-Gen-P & 56\,{\tiny$\pm$11} & 68\,{\tiny$\pm$4} & 45\,{\tiny$\pm$7} & 49\,{\tiny$\pm$6} & 24 & 13 & 32 & 13 & $\cdot$ \\
\quad KL-Gen & 184\,{\tiny$\pm$19} & 98\,{\tiny$\pm$3} & 128\,{\tiny$\pm$12} & 51\,{\tiny$\pm$6} & 30 & $\cdot$ & $\cdot$ & $\cdot$ & $\cdot$ \\
\midrule
\textit{Oracle (reference)} & 9\,{\tiny$\pm$2} & 37\,{\tiny$\pm$2} & 18\,{\tiny$\pm$1} & 14\,{\tiny$\pm$2} & 3 &  &  &  &  \\
\bottomrule
\end{tabular}}

\end{table}

\section{Capacity Expansion} \label{app:cep}

\subsection{Model} \label{app:cep_model}

\paragraph{Variables.} Let $\mathcal Z$ be the 20 zones, indexed by $n$, $\mathcal{L}$ the candidate lines with zone-line incidence matrix $A$, $\mathcal{G}$ the technologies, of which $\mathcal G_{\rm th}$ are gas and $\mathcal G_{\rm cont} = \mathcal G \setminus \mathcal G_{\rm th}$ the others, $\mathcal{S}$ the scenarios, and $t = 1, \dots, 24$ the hours of a day. The first stage builds generation capacity $x^{\rm gen}_{ng}$ and transmission capacity $x^{\rm tr}_{\ell}$ from zero. Gas capacity is $\bar P_g u_{ng}$ for a plant count $u_{ng}$, which is relaxed to be continuous throughout. The recourse variables of scenario $s$, collected in $y_s$, are dispatch $p_{stng}$, flows $f_{st\ell}$, load shedding $\ell^{\rm shed}_{stn}$, and battery charge $p^{\rm ch}_{stn}$, discharge $p^{\rm dis}_{stn}$, and state of charge $e_{stn}$.

\paragraph{Objective.} Given scenario weights $p_s$ summing to one (App.~\ref{app:optimization}), the master is
\begin{align}
\min \quad & C^{\rm inv}(x) + (1-\beta) \sum_{s\in\mathcal{S}} p_s\, \nu\, C^{\rm op}_s(x,y_s) + \beta \Big( \zeta + \frac{1}{1-\alpha} \sum_{s\in\mathcal{S}} p_s\, v_s \Big) \label{eq:cep_objective} \\
\text{s.t.} \quad & v_s \geq \nu\, C^{\rm op}_s(x,y_s) - \zeta, \quad v_s \geq 0, \quad \forall s, \qquad C^{\rm inv}(x) \leq B, \nonumber
\end{align}
together with the constraints below, where $\nu = 8760/24$ annualizes a day's operating cost, $B$ is the capital budget below, $\beta = 0.25$, $\alpha = 0.9$, and $\zeta$ and $v_s$ are the auxiliary variables of the Rockafellar--Uryasev representation of CVaR \citep{rockafellar2000optimization}. The investment and operating costs are
\begin{align*}
C^{\rm inv}(x) &= \sum_{n,\, g\in\mathcal{G}_{\rm cont}} c^{\rm inv}_g x^{\rm gen}_{ng} + \sum_{n,\, g\in\mathcal{G}_{\rm th}} c^{\rm inv}_g \bar P_g u_{ng} + \sum_{\ell\in\mathcal{L}} c^{\rm tr} |\ell|\, x^{\rm tr}_{\ell}, \\
C^{\rm op}_s(x,y_s) &= \sum_{t,n,g} c^{\rm var}_g p_{stng} + c^{\rm shed}\sum_{t,n} \ell^{\rm shed}_{stn},
\end{align*}
where $|\ell|$ is the length of line $\ell$, $c^{\rm inv}_g$ and $c^{\rm tr}$ are annualized, and for gas $c^{\rm var}_g = c^{\rm fuel}\,h_g + c^{\rm vom}_g + c^{\rm CO2} r^{\rm CO2} h_g$, with variable operation and maintenance cost $c^{\rm vom}_g$, heat rate $h_g$, and emissions rate $r^{\rm CO2}$.

\paragraph{Constraints.} For all $s$, $t$, and $n$, zonal balance is
\begin{align*}
\sum_g p_{stng} + p^{\rm dis}_{stn} - p^{\rm ch}_{stn} + \ell^{\rm shed}_{stn} - d_{stn} = (A f_{st})_n,
\end{align*}
where $d_{stn}$ is the scenario's demand in zone $n$ at hour $t$, and flows are limited by $-x^{\rm tr}_{\ell} \le f_{st\ell} \le x^{\rm tr}_{\ell}$, a network-flow model without power-flow constraints. Wind and solar dispatch is bounded by the scenario's capacity factors, $p_{stng} \le a_{stng}x^{\rm gen}_{ng}$, and gas dispatch by installed capacity, $p_{stng} \le \bar P_g u_{ng}$. For batteries,
\begin{align*}
0 \le p^{\rm ch}_{stn},p^{\rm dis}_{stn} \le x^{\rm gen}_{n,{\rm bat}}, \qquad
0 \le e_{stn} \le H x^{\rm gen}_{n,{\rm bat}}, \qquad
e_{stn} = e_{s,t-1,n} + \eta^{\rm ch}p^{\rm ch}_{stn} - p^{\rm dis}_{stn}/\eta^{\rm dis},
\end{align*}
where $x^{\rm gen}_{n,{\rm bat}}$ is the battery power capacity of zone $n$, with the last hour linked cyclically to the first. Load shedding satisfies $0 \le \ell^{\rm shed}_{stn} \le \max\{d_{stn},0\}$. Gas plants follow a linearized clustered unit commitment: the committed count $c_{stng} \in [0, u_{ng}]$ of each gas technology in each zone changes by start-ups and shut-downs, output satisfies $\underline{p}_g \bar P_g c_{stng} \le p_{stng} \le \bar P_g c_{stng}$, started and stopped units respect minimum up- and down-times (cyclic over the day), and each start-up adds $c^{\rm su}_g \bar P_g$ to $C^{\rm op}_s$. The count $c_{stng}$ is continuous, so the recourse remains a linear program. There are no ramping constraints.

\paragraph{Capital budget.} $B = 21.5$~bn\$/yr for every method, radius, record, seed, and $k$. It was fixed before any run, inside the range (19.5 to 23.6~bn\$/yr) of the investment cost of unbudgeted plans from a preliminary configuration of KL-Rec at $k = 1$ ($\rho = 0$, $\rho_{\mathrm{ref}} = 0.4$). At $B = 19.35$ and $23.65$~bn\$/yr ($\pm 10\,\%$), KL-Gen remains cheaper out of distribution than KL-Rec and KL-OT at $(\rho_{\mathrm{ref}}, \rho) = (0.2, 0.05)$ at every $k$.

\begin{figure}[htbp]
    \centering
    \includegraphics[width=0.6\linewidth]{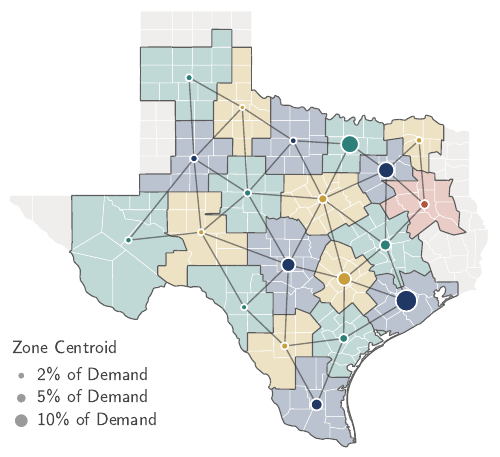}
    \caption{The 20 zones of the capacity expansion model. Counties of the ERCOT region are shaded by zone (other Texas counties in gray) and zone boundaries are drawn in dark gray. Lines join zone centroids along the 40 candidate transmission corridors, and centroid markers are sized by the zone's share of 2010 demand.}
    \label{fig:cep_zones}
\end{figure}

\paragraph{Network and parameters.} The 214 counties of the region of the Electric Reliability Council of Texas (ERCOT) are grouped into 20 zones by $k$-means on county locations. Dallas and Rockwall counties, which have no coordinates in the source table, take their 2020 Census internal points. Zonal demand is the sum of county demand, and zonal capacity factors are county averages. Candidate lines join zone centroids along the edges of their Gabriel graph (40 lines, 116 to 279~km, median 178~km). Figure~\ref{fig:cep_zones} shows the zones and lines, and Table~\ref{tab:tech_params} gives the cost and technology parameters.

\begin{table}[htbp]
\centering
\caption{Cost and technology parameters, in real 2024 dollars. Generation costs are from the 2024 Annual Technology Baseline of the National Renewable Energy Laboratory \citep{mirletz_2024_2024} (Moderate scenario, 2030, converted from 2022 dollars by $\times 1.061$), and heat rates are assumed. The gas price and emissions rate are from the Energy Information Administration (Annual Energy Outlook 2024, 2030). Capital costs are annualized with the capital recovery factor $\mathrm{CRF}(r,L)=r(1+r)^L/((1+r)^L-1)$ over the lifetime $L$, plus fixed operation and maintenance (O\&M). VOM is variable O\&M.}
\label{tab:tech_params}
\small
\begin{tabular}{lrrrrr}
\toprule
Technology & Capital & Fixed O\&M & VOM & Heat rate & Lifetime \\
 & (\$/kW) & (\$/kW-yr) & (\$/MWh) & (MMBtu/MWh) & (yr) \\
\midrule
Wind        & 1494 & 31.0 & 0.0 & --  & 30 \\
Solar       & 1266 & 19.1 & 0.0 & --  & 30 \\
Gas, combined cycle & 1489 & 33.7 & 2.2 & 6.4 & 30 \\
Gas, combustion turbine & 1348 & 26.5 & 7.4 & 9.5 & 30 \\
Battery 4h  & 1540 & 34.5 & 0.0 & --  & 15 \\
\bottomrule
\end{tabular}

\vspace{6pt}
\begin{tabular}{llll}
\toprule
Parameter & Value & Parameter & Value \\
\midrule
Lost load $c^{\rm shed}$ & \$10{,}000/MWh & Discount rate $r$ & 7\% \\
Carbon price $c^{\rm CO2}$ & \$150/tCO$_2$ & Gas units (combined cycle, turbine) & 500, 150 MW \\
Gas price $c^{\rm fuel}$ & \$3.43/MMBtu & Battery duration $H$ & 4 h \\
Emissions rate $r^{\rm CO2}$ & 0.05306 t/MMBtu & Efficiency $\eta^{\rm ch} = \eta^{\rm dis}$ & $\sqrt{0.85}$ \\
Transmission $c^{\rm tr}$ & \$1{,}500/MW-km & Line lifetime, fixed O\&M & 40 yr, 2\%/yr \\
Minimum output $\underline p_g$ (CC, CT) & 40\%, 30\% & Minimum up/down time (CC, CT) & 6 h, 1 h \\
Start-up cost $c^{\rm su}_g$ (CC, CT) & \$79, \$179/MW & & \\
\bottomrule
\end{tabular}
\end{table}

\subsection{Data and Records} \label{app:cep_data}

\paragraph{Data.} The source is hourly county-level ERCOT data: wind and solar capacity factors from \citet{bracken2023hourly} and demand simulated by the Total ELectricity Load (TELL) model \citep{mcgrath2022tell}, aggregated to the 20 zones. A scenario is one day of 20 zonal demands and 40 zonal capacity factors, so $D = 1{,}440$, cut at local midnight. Short gaps are interpolated, and days with longer gaps are dropped. Table~\ref{tab:splits} gives the periods.

\begin{table}[htbp]
\centering
\caption{Data periods. Records A to J are the years 2010, 2009, 2007, 2008, 2005, 2001, 2002, 2003, 2004, and 2006. The years 2015--2019 are used only to place 2020 among the other years.}
\label{tab:splits}
\small
\begin{tabular}{llrl}
\toprule
Period & Years & Days & Role \\
\midrule
Record & 2001--2010 & $10 \times 364$ & records and fitted nominals \\
In distribution (ID) & 2011--2014 & 1{,}458 & evaluation \\
Out of distribution (OOD) & 2020 & 364 & evaluation \\
\bottomrule
\end{tabular}
\end{table}

\paragraph{Records.} At $k = 1$ a record is one year, at $k = \tfrac12$ every second day of it (182 days), and at $k = 5$ one of five sets of five years. Every record is run with five seeds.

\paragraph{Out-of-distribution year.} 2020 was designated out of distribution before any method was scored on it, although its days later also served for development (App.~\ref{app:cep_eval}). It is the most extreme year of 2001--2020 in the tail of demand: its annual peak (77.4 GW) lies 2.5 standard deviations of year-to-year variation above the 2001--2019 trend. Net load is demand less the output of a fixed reference fleet of 62.6 GW of wind and 24.3 GW of solar, the mean zonal capacities of 25 KL-OT plans at $k = \tfrac12$. The 99th percentile of daily peak net load is 50.4 GW in 2020, against at most 46.7 GW in any other year.

\subsection{Learned Nominals} \label{app:generator}

KL-Gen and KL-Gen-P use a conditional normalizing flow, and KL-Gen (copula) and KL-Gen-P (copula) a Gaussian copula with a separate mean and covariance for each season. Both are fitted to the record alone and share their marginal transform and smoothing.

\paragraph{Copula.} Each coordinate of a day, in log demand and logit capacity factors, is mapped to its normal score among the record's days, and the scores of each season define a Gaussian. A draw uses one extra latent coordinate $z_0$ to choose the season in proportion to the record's season counts, forms the seasonal Gaussian draw from the other coordinates of $z$, maps each coordinate back through the record's empirical quantiles, and smooths the day over adjacent hours with weights $(0.2, 0.6, 0.2)$.

\paragraph{Flow.} The flow replaces the seasonal Gaussian with a model of the 16 leading whitened principal components of the scores, a conditional RealNVP \citep{dinh_density_2017} with an affine layer and two coupling layers of width 16 (2{,}720 parameters), conditioned on two harmonics of the day of the year and on the season. The remaining components follow a seasonal Gaussian. The flow is fitted by maximum likelihood with AdamW, with Gaussian noise of standard deviation 0.3 added to the components, for a number of steps chosen on a season-stratified 20\% of the record's days, and then refitted on all of them. All seeds of a record share one flow. In a draw, $z_0$ selects a record day, whose calendar day and season are the covariates, and the other coordinates drive the flow and the Gaussian remainder. The day is then decoded and smoothed as for the copula.

\paragraph{Properties and fidelity.} Before smoothing, each coordinate of a draw of either nominal lies within the record's range, although a draw can combine coordinates into a system-wide peak that no record day reaches. In a separate study on 2001--2010, the flow's held-out log-likelihood exceeded that of a seasonal Gaussian of the copula's form, with tuned shrinkage, in each of the ten years, by a median of 0.97 nats per day with 182 training days and 1.85 with 320 (three training seeds). Its configuration was fixed from that study before any plan was scored. The most severe scenarios KL-Gen admits are less spatially coherent in solar than real days (mean correlation of zonal solar profiles 0.82, against 0.87 for the 20 most severe days of 2020). Both nominals overstate the frequency of the decade's tail: at $k = \tfrac12$, the share of their draws whose peak net load exceeds the 2001--2010 99th percentile is about 1.7 times the decade's own share for the copula and 2.1 times for the flow.

\subsection{Methods} \label{app:cep_method}

Table~\ref{tab:cep_methods} summarizes the methods. KL-Rec, KL-Gen, KL-Gen-P, their copula variants, and KL-OT share the adaptive loop of Algorithm~\ref{alg:etas}:
\begin{description}[leftmargin=1em, itemsep=1pt, topsep=2pt]
    \item[Rounds.] The initial plan solves the unbudgeted, risk-neutral problem on one scenario fixed by the seed. Each of twelve rounds evaluates $M = 256$ candidates at the current plan (for the proposal methods, an initial sample of $M = 64$, to which the proposal is fitted, and 192 draws from it), tilts them at $\rho_{\mathrm{ref}}$, and admits $n = 4$, so the master holds at most 48 scenarios.
    \item[Base weights.] A new scenario's base weight, $c_i/(M\tau_i)$, or $c_i b_i/\tau_i$ for the proposal methods, where $b_i \propto \varphi(z_i)/r_t(z_i)$ and $\tau_i \propto b_i \exp(Q_i/\lambda_t)$, is fixed at admission, and the weights of all rounds are normalized together.
    \item[Master.] Each round recomputes the worst case at the current plan (App.~\ref{app:optimization}) and solves for the next plan. The last plan is reported.
    \item[Radii.] KL-Rec, KL-OT, and the copula variants are run at $k \le 1$ at the eleven pairs with $\rho_{\mathrm{ref}} \in \{0.2, 0.4\}$, $\rho \in \{0, 0.02, 0.05, 0.1, 0.2, 0.4\}$, and $\rho \le \rho_{\mathrm{ref}}$. KL-Gen, KL-Gen-P, and every method at $k = 5$ are run at the five pairs $(0.2, 0)$, $(0.2, 0.05)$, $(0.2, 0.2)$, $(0.4, 0)$, and $(0.4, 0.05)$.
\end{description}
The methods differ as follows.
\begin{description}[leftmargin=1em, itemsep=1pt, topsep=2pt]
    \item[KL-Rec.] Candidates are record days drawn with replacement.
    \item[KL-Gen.] Candidates are draws of the flow.
    \item[KL-Gen-P.] Candidates are 192 draws from a proposal that shifts the mean of $\varphi$ (the mean-shift case of Sec.~\ref{sec:etas}) in a 10-dimensional active subspace \citep{constantine2015active} of the cost gradients of 64 fitting draws, mixed with $\varphi$ at $a = 0.5$. Four are resampled by the tilt.
    \item[KL-Gen (copula) and KL-Gen-P (copula).] KL-Gen and KL-Gen-P with the copula in place of the flow, with all else identical. The tables report KL-Gen (copula).
    \item[KL-OT ($\delta$).] Candidates are 256 record days. From the second round, each is moved before evaluation to an approximate maximizer of $Q(x_t, \xi) - \gamma \|\omega(\xi) - \omega(\xi^0)\|_2^2$, the Lagrangian of a type-2 Wasserstein ball, where $\xi^0$ is the record day and $\omega$ its standardized log demand and logit capacity factors, by three damped fixed-point steps on the dispatch subgradient. A moved day keeps its record day's base weight. The multiplier $\gamma$ is set once, at the second round, so that the mean cost increase of moved days is $\delta$ times the mean cost of the unmoved candidates. The transport budget $\delta \in \{0.3, 1, 3\}$ is thus a cost-increase target at one plan rather than a radius.
    \item[W-DRO ($\delta$).] Holds 48 record days with uniform weights and no KL ball. Each round moves every day by the step of KL-OT, with $\gamma$ set as for KL-OT, to the target $\delta$, at the first plan, and re-solves the master on each day's largest penalized cost, for at most four rounds. At $\delta = 1$, 68 of 100 runs reach the four-round cap, with a median relative gap of 2 to $4 \times 10^{-3}$ across $k$.
    \item[CSSC.] Selects 48 days from 256 record days (all 182 at $k = \tfrac12$) by the clustering program of \citet{keutchayan_problem-driven_2023}, with opportunity costs $C^{\rm inv}(x^*_i) + \nu\, Q(x^*_i, \xi_j)$ for the plan $x^*_i$ that solves the problem on day $i$ alone, and weights them by cluster mass. It has no KL ball.
    \item[SAA.] Solves the master on 48 record days with uniform weights.
\end{description}

\begin{table}[htbp]
\centering
\caption{Capacity expansion methods. All use the objective in \eqref{eq:cep_objective} with $\beta = 0.25$, $\alpha = 0.9$, and the budget $B$.}
\label{tab:cep_methods}
\small
\begin{tabular}{lp{0.28\linewidth}p{0.22\linewidth}p{0.18\linewidth}}
\toprule
Method & Candidates per round & Scenarios in the master & Weights \\
\midrule
KL-Rec & 256 record days & at most 48, four per round & KL worst case at $\rho$ \\
KL-Gen & 256 draws of the flow & same & same \\
KL-Gen-P & 192 draws from a fitted proposal & same & same \\
KL-Gen (copula) & 256 copula draws & same & same \\
KL-Gen-P (copula) & 192 draws from a proposal fitted to the copula & same & same \\
KL-OT ($\delta$) & 256 record days, moved from the second round & same & same \\
W-DRO ($\delta$) & a moved copy of each day, from the second round & 48 record days, each valued at its largest penalized cost over the day and its moved copies & $1/48$ \\
CSSC & 256 record days (all 182 at $k = \tfrac12$) & 48 selected & cluster masses \\
SAA & none & 48 record days & $1/48$ \\
\bottomrule
\end{tabular}
\end{table}

\subsection{Solving the Master} \label{app:optimization}

\paragraph{The worst case.} For the mean--CVaR risk measure, with $Q_i \coloneqq Q(x, \xi_i)$ and normalized base weights $\bar w$, the adversary maximizes $\min_\zeta \{ \zeta + \sum_i p_i\, \psi_\zeta(Q_i) \}$ over the ball $\mathcal B_\rho$ of probability vectors $p$ with $\sum_i p_i \log(p_i/\bar w_i) \leq \rho$. The objective is linear in $p$ and convex in $\zeta$ and the feasible set is convex and compact, so the maximum and minimum may be exchanged \citep{sion1958general}. For fixed $\zeta$ the maximizer is the tilt $p_i \propto \bar w_i \exp(\psi_\zeta(Q_i)/\lambda)$, with $\lambda$ bisected so that the KL constraint binds, and $\zeta$ is found by golden-section search over $[\min_i Q_i, \max_i Q_i]$.

\paragraph{The alternation.} Each round solves the linear program \eqref{eq:cep_objective} with the weights fixed at the worst case computed at the current plan, so the reported plan $x_T$ is a best response to the worst case at $x_{T-1}$ rather than a solution of the KL-DRO problem on the final scenario set. Weak duality bounds its optimality gap. With $J(x, p)$ the value of \eqref{eq:cep_objective} at plan $x$ and weights $p$, minimized over the recourse and auxiliary variables, $p_T$ the final weights, and $p'$ the worst case at $x_T$, the gap is at most $\max_{p \in \mathcal B_\rho} J(x_T, p) - \max\{\min_x J(x, p_T), \min_x J(x, p')\}$, which costs one more master solve. Over 200 runs of KL-Rec, KL-OT, KL-Gen, and KL-Gen (copula) at $(\rho_{\mathrm{ref}}, \rho) = (0.2, 0.05)$ and $k = \tfrac12$, the bound is small for a typical run (median 0.02 to 0.21 bn\$/yr by method, under 0.7\% of the objective) but not for every run (90th percentile 0.4 to 1.1 bn\$/yr, and above 0.5 bn\$/yr in 36 runs). Individual plans are therefore not settled, and one further step moves a run's OOD cost by more than 0.5 bn\$/yr in 100 of the 200 runs. These moves are small against the variation over seeds (median 0.5 against a seed standard deviation of 4.1 bn\$/yr), and they leave the comparisons intact. Re-scored at the further step, every comparison of KL-Gen with KL-Rec, KL-OT, and KL-Gen (copula) at this radius pair keeps its sign, in and out of distribution, and KL-Gen's out-of-distribution advantage is 15.4~bn\$/yr over KL-Rec and 2.5 over KL-OT, against 18.3 and 2.4 at the reported plans.

\paragraph{Solver and computation.} Masters are built directly in the Gurobi Python interface and solved by the Gurobi barrier method \citep{gurobi} on 8 threads, and the candidates' recourse costs are evaluated in parallel on 8 cores.

\subsection{Evaluation and Statistics} \label{app:cep_eval}

\paragraph{Scoring.} The final plan of every run is scored by the objective in \eqref{eq:cep_objective} with uniform weights over all ID or OOD days.

\paragraph{Statistics.} Paired differences between methods at the same seed are summarized by their median over the five seeds, so the record is the unit of analysis: ten records at $k = \tfrac12$, five (A to E) at $k = 1$, and five five-year blocks at $k = 5$. Table~\ref{tab:cep-paired} gives the median over records with its standard error, the standard deviation over records divided by the square root of their number, and a two-sided Wilcoxon signed-rank test.

\paragraph{Development.} The radii, the risk weight $\beta$, the marginals and smoothing of the nominals, and the construction of the base weights were chosen after comparing configurations on the ID and OOD days, which therefore served for development as well as for evaluation. The flow's architecture and training were fixed before any plan was scored (App.~\ref{app:generator}), and it shares the copula's marginals and smoothing.

\paragraph{Severity.} A day's severity is its system peak net load, with net load as in App.~\ref{app:cep_data}, and its peak hour is the hour of that peak. The OOD days of Figure~\ref{fig:cep} are the 20 most severe days of 2020 (42.2 to 54.3 GW), and a method's most severe scenarios are the top 10\% by severity of the scenarios it admits, 48 per run, pooled over record years and seeds at $(\rho_{\mathrm{ref}}, \rho) = (0.2, 0.05)$.

\paragraph{Figure~\ref{fig:cep}(a).} Each point is a setting's median over records of the median over seeds, and a method's frontier joins its non-dominated settings. KL-OT at $\delta = 3$ was run at three of the five radius pairs, and W-DRO is a single point. Further radius pairs of KL-Rec and KL-OT are dominated by the KL-Gen points shown, except KL-OT at $(0.2, 0.02)$, which is cheaper in distribution than every KL-Gen setting.

\paragraph{Figure~\ref{fig:cep}(b).} The maps show zonal wind capacity factors, standardized per zone over the day and clipped at $\pm 2$, on the most severe day of 2020 and on each method's most severe admitted scenario over all 50 runs. Spatial coherence is the mean correlation of the 24-hour profiles over the 190 pairs of zones, over daylight hours for solar. Boxes span the interquartile range, whiskers the 10th to 90th percentiles, and the gray band is the interquartile range over the 20 OOD days.

\paragraph{Figure~\ref{fig:cep}(c).} Each day is represented by its 20 zonal net loads at its peak hour, standardized per zone over 2001--2010. The horizontal axis is peak net load relative to the median OOD day. The vertical axis is the distance to the nearest real day of 2001--2010 outside the record's year, after the scenario's demand is scaled so that its system net load equals that day's, so only the spatial shape counts. Small markers are medians per record, large markers medians over records, the star the median OOD day, and the dotted lines the OOD days' median severity and the median distance of the decade's own most severe days. A small distance marks a shape common among real days but does not prove realism. Over the full day (24 hours of demand, wind, and solar), the most severe scenarios of KL-Gen lie 0.424 from the nearest real day of 2011--2014 and 2020, against 0.367 for real days of the same severity and 0.372 for KL-OT ($\delta = 1$). Compared with each OOD day, the nearest admitted scenarios of KL-Gen lie within 0.2 of the OOD days' demand level, against 0.5 to 1.0 below for KL-OT and KL-Rec. With demand rescaled, the three methods' shape distances are nearly the same.

\paragraph{Figure~\ref{fig:cep}(d).} Demand is system demand at the scenario's peak hour, relative to the largest such value over the 2001--2010 days (72.6 GW), and wind is system wind output at that hour under the reference fleet. Gray points are the 2001--2010 days; some of them, and seven scenarios of KL-OT ($\delta = 3$), lie outside the axes.

\paragraph{Reported settings.} KL-OT and W-DRO were also run at $\delta \in \{0.3, 3\}$, KL-OT at four radius pairs for $k \le 1$ and two for $k = 5$. These settings never reach the Pareto front, so the tables report $\delta = 1$, and Figure~\ref{fig:cep} also shows KL-OT at $\delta = 3$.

\subsection{Full Results} \label{app:cep_results}

\begin{figure}[htbp]
    \centering
    \includegraphics[width=\linewidth]{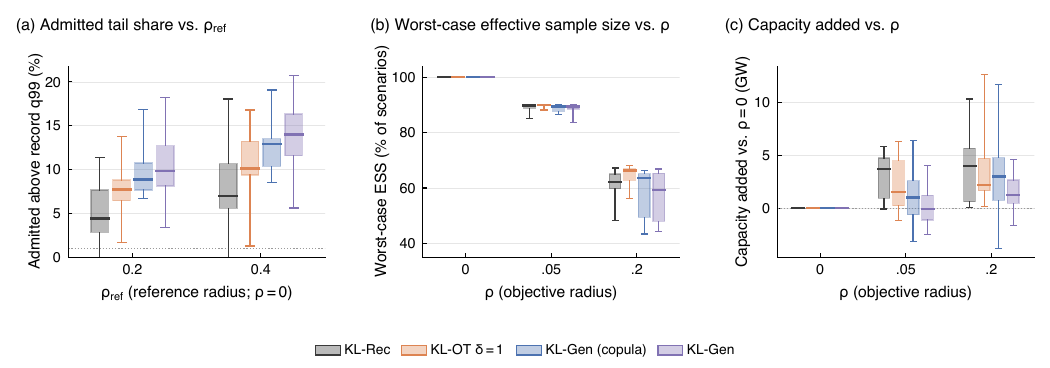}
    \caption{The two radii act as designed (capacity expansion, $k = \tfrac12$). Boxes span the interquartile range over the ten records (each the median over five seeds; in (a), seeds are pooled), and whiskers the full range. \textbf{(a)} Share of admitted scenarios above the record's 99th percentile of peak net load, at $\rho = 0$ (dotted: the record's own 1\,\%). Raising $\rho_{\mathrm{ref}}$ admits more severe scenarios, in 37 of 40 (method, record) pairs. \textbf{(b)} Effective sample size of the worst-case weights, as a share of the admitted scenarios, at $\rho_{\mathrm{ref}} = 0.2$. It falls as $\rho$ grows, and the realized divergence equals $\rho$ in every run. \textbf{(c)} Capacity added relative to $\rho = 0$ (GW) at $\rho_{\mathrm{ref}} = 0.2$, paired within record and seed. Raising $\rho$ moves capital toward the budget and changes the mix.}
    \label{fig:cep-dials}
\end{figure}

\begin{figure}[htbp]
    \centering
    \includegraphics[width=\linewidth]{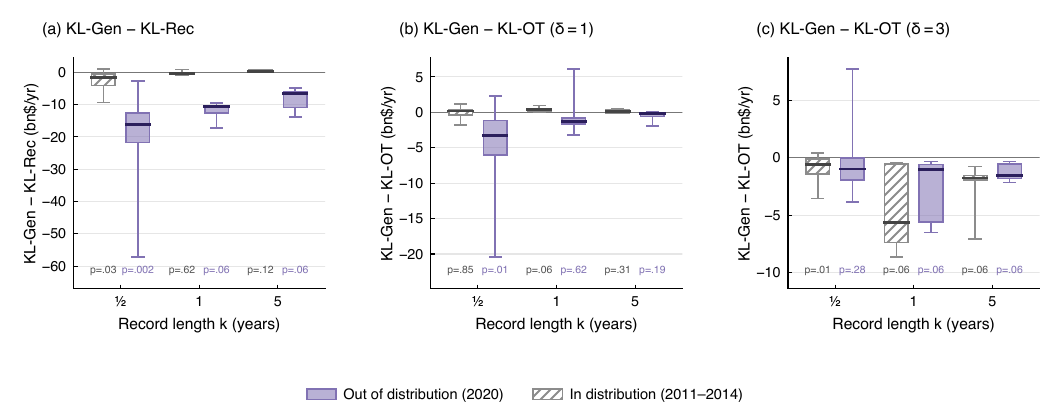}
    \caption{KL-Gen's advantage across record lengths (capacity expansion). Paired cost difference (bn\$/yr; negative: KL-Gen is cheaper) against \textbf{(a)} KL-Rec, \textbf{(b)} KL-OT ($\delta = 1$), and \textbf{(c)} KL-OT ($\delta = 3$), out of distribution (filled) and in distribution (hatched). Each record contributes the median, over the radius pairs of Table~\ref{tab:cep-pareto} at which both methods were run (all five; for $\delta = 3$, three at $k \le 1$ and two at $k = 5$), of its median over seeds. Boxes span the interquartile range over records and whiskers the full range; the numbers are two-sided Wilcoxon signed-rank $p$-values over records (with five records, the smallest attainable is 0.0625). The out-of-distribution gap to KL-Rec and to KL-OT ($\delta = 1$) shrinks as the record grows; KL-OT ($\delta = 3$) narrows it at $k = \tfrac12$ at a higher in-distribution cost.}
    \label{fig:cep-kcross}
\end{figure}

Tables~\ref{tab:cep-pareto} and~\ref{tab:cep-paired} give the full results, in which the copula ablation is the two KL-Gen (copula) columns of Table~\ref{tab:cep-pareto} and the two KL-Gen $-$ KL-Gen (copula) columns of Table~\ref{tab:cep-paired}. Figure~\ref{fig:cep-dials} shows that the two radii act as designed, and Figure~\ref{fig:cep-kcross} summarizes KL-Gen's advantage across record lengths.

\paragraph{Paired comparisons.} Table~\ref{tab:cep-paired} compares methods at the same radius pair $(\rho_{\mathrm{ref}}, \rho)$. With five records at $k = 1$ and five blocks at $k = 5$, no difference can reach $p < 0.05$, so we report the number of records in which the first method is cheaper.

\emph{KL-Gen vs.\ KL-Rec.} KL-Gen's out-of-distribution cost is lower at every radius pair:
\begin{itemize}[leftmargin=1.5em, itemsep=1pt, topsep=2pt]
  \item at $k = \tfrac12$, by 12.7--22.2~bn\$/yr, in 10 of 10 records ($p < 0.01$ at all five pairs);
  \item at $k = 1$, by 8.4--14.5~bn\$/yr, in 5 of 5 records;
  \item at $k = 5$, by 3.8--13.8~bn\$/yr, in 5 of 5 blocks.
\end{itemize}
The gap shrinks as the record grows in the median over radius pairs (Figure~\ref{fig:cep-kcross}), though not at every pair. In distribution, KL-Gen is cheaper at $k = \tfrac12$ (by 0.46--2.92~bn\$/yr), mixed at $k = 1$, and up to 0.73~bn\$/yr more expensive at $k = 5$.

\emph{KL-Gen vs.\ KL-OT.} KL-Gen's out-of-distribution cost is lower at all 15 radius pairs:
\begin{itemize}[leftmargin=1.5em, itemsep=1pt, topsep=2pt]
  \item at $k = \tfrac12$, by 2.3--6.5~bn\$/yr, in 7--9 of 10 records ($p < 0.05$ at four pairs);
  \item at $k = 1$, by 0.4--1.9~bn\$/yr, in 3--5 of 5 records;
  \item at $k = 5$, by 0.07--0.76~bn\$/yr, in 3--4 of 5 blocks.
\end{itemize}
In distribution the two are indistinguishable at $k = \tfrac12$. At $k \ge 1$, KL-Gen costs up to 0.45~bn\$/yr more, and is cheaper in at most 3 of 5 records. At $k \ge 1$ and the same radii, KL-Gen therefore trades a small in-distribution premium for lower out-of-distribution cost.

\emph{Ablations.} The proposal variant stays within 0.65~bn\$/yr of KL-Gen on both axes, and none of its ten differences at $k = \tfrac12$ is significant. Out of distribution, KL-Gen and KL-Gen (copula) are within two standard errors of each other at every radius pair and $k$ except two pairs at $k = 5$, where KL-Gen is cheaper (by 0.31 and 0.84~bn\$/yr). In distribution, the copula is up to 0.36~bn\$/yr cheaper at $k \ge 1$, beyond two standard errors at one pair at $k = 1$ and two at $k = 5$. Replacing the copula with the flow thus changes out-of-distribution cost little and adds a small in-distribution premium when the record is longer.

\begin{table}[t]
\centering
\scriptsize
\setlength{\tabcolsep}{3.2pt}
\caption{Capacity expansion: in-distribution (2011--2014) and out-of-distribution (2020) cost (bn\$/yr) at the five radius pairs $(\rho_{\mathrm{ref}}, \rho)$ run at every $k$, and of SAA, CSSC, and W-DRO, which have no KL radii (median over records of the median over five seeds). Standard errors over records are 0.02--1.6 in distribution and 0.05--6.5 out of distribution for the methods with radii, and up to 6.7 and 15.6 for the others. Table~\ref{tab:cep-paired} gives paired differences. Bold: on the Pareto front at that $k$. KL-OT and W-DRO use $\delta = 1$.}
\label{tab:cep-pareto}
\begin{tabular}{cccccccccccc}
\toprule
 & & \multicolumn{5}{c}{In-distribution} & \multicolumn{5}{c}{Out-of-distribution} \\
\cmidrule(lr){3-7}\cmidrule(lr){8-12}
$\rho_{\mathrm{ref}}$ & $\rho$ & KL-Rec & KL-OT & KL-Gen & KL-Gen-P & KL-Gen (copula) & KL-Rec & KL-OT & KL-Gen & KL-Gen-P & KL-Gen (copula) \\
\midrule
\multicolumn{12}{l}{\textit{$k = \frac12$ (10 record years)}} \\
0.2 & 0 & 33.44 & 30.35 & \textbf{30.34} & 30.71 & 30.39 & 56.96 & 38.42 & \textbf{33.91} & 33.35 & 38.79 \\
 & 0.05 & 32.67 & 30.43 & 30.89 & \textbf{30.65} & 31.09 & 53.62 & 35.97 & 33.07 & \textbf{32.90} & 33.51 \\
 & 0.2 & 32.44 & 30.73 & 31.14 & 31.00 & 31.08 & 50.19 & 38.86 & 33.19 & 33.17 & 33.66 \\
\cmidrule(lr){1-12}
0.4 & 0 & 32.31 & \textbf{30.29} & \textbf{30.53} & \textbf{30.48} & 30.34 & 55.71 & \textbf{40.57} & \textbf{33.58} & \textbf{33.60} & 34.30 \\
 & 0.05 & 31.70 & 30.73 & 30.88 & 31.00 & \textbf{30.66} & 47.46 & 37.35 & 33.08 & 32.97 & \textbf{32.88} \\
\cmidrule(lr){1-12}
\multicolumn{2}{l}{SAA} & \multicolumn{5}{c}{42.79} & \multicolumn{5}{c}{97.06} \\
\multicolumn{2}{l}{CSSC} & \multicolumn{5}{c}{34.12} & \multicolumn{5}{c}{49.10} \\
\multicolumn{2}{l}{W-DRO} & \multicolumn{5}{c}{31.42} & \multicolumn{5}{c}{41.42} \\
\midrule
\multicolumn{12}{l}{\textit{$k = 1$ (5 record years)}} \\
0.2 & 0 & 30.83 & 30.27 & \textbf{30.34} & 30.41 & 30.45 & 49.67 & 35.98 & \textbf{32.87} & 33.11 & 34.83 \\
 & 0.05 & 30.56 & 30.49 & 30.75 & \textbf{31.14} & 31.07 & 43.43 & 34.46 & 33.02 & \textbf{32.77} & 33.11 \\
 & 0.2 & 30.82 & 30.96 & 31.08 & 31.08 & 31.32 & 42.76 & 34.16 & 33.21 & 33.18 & 33.29 \\
\cmidrule(lr){1-12}
0.4 & 0 & 30.36 & \textbf{30.02} & \textbf{30.20} & 30.44 & 30.26 & 46.48 & \textbf{35.70} & \textbf{32.96} & 32.88 & 33.10 \\
 & 0.05 & 30.58 & 30.41 & 30.91 & 31.13 & \textbf{30.69} & 46.60 & 33.67 & 32.85 & 33.02 & \textbf{32.79} \\
\cmidrule(lr){1-12}
\multicolumn{2}{l}{SAA} & \multicolumn{5}{c}{40.99} & \multicolumn{5}{c}{83.64} \\
\multicolumn{2}{l}{CSSC} & \multicolumn{5}{c}{33.65} & \multicolumn{5}{c}{45.32} \\
\multicolumn{2}{l}{W-DRO} & \multicolumn{5}{c}{30.69} & \multicolumn{5}{c}{36.43} \\
\midrule
\multicolumn{12}{l}{\textit{$k = 5$ (5 record blocks)}} \\
0.2 & 0 & 30.11 & 30.18 & 30.05 & \textbf{30.24} & \textbf{29.98} & 42.59 & 34.77 & 33.54 & \textbf{32.65} & \textbf{33.97} \\
 & 0.05 & 30.73 & 30.53 & 30.80 & 30.60 & 30.52 & 46.28 & 33.26 & 32.72 & 32.81 & 33.16 \\
 & 0.2 & 30.83 & 30.71 & 31.02 & 30.68 & 31.02 & 44.18 & 34.69 & 33.27 & 33.13 & 33.58 \\
\cmidrule(lr){1-12}
0.4 & 0 & \textbf{29.78} & \textbf{29.89} & 30.37 & \textbf{30.19} & \textbf{30.03} & \textbf{39.75} & \textbf{35.24} & 32.76 & \textbf{33.05} & \textbf{33.34} \\
 & 0.05 & \textbf{29.88} & 30.29 & \textbf{30.87} & 30.92 & 30.54 & \textbf{37.09} & 33.12 & \textbf{32.61} & 32.62 & 33.04 \\
\cmidrule(lr){1-12}
\multicolumn{2}{l}{SAA} & \multicolumn{5}{c}{42.97} & \multicolumn{5}{c}{88.00} \\
\multicolumn{2}{l}{CSSC} & \multicolumn{5}{c}{32.73} & \multicolumn{5}{c}{42.12} \\
\multicolumn{2}{l}{W-DRO} & \multicolumn{5}{c}{34.78} & \multicolumn{5}{c}{38.51} \\
\bottomrule
\end{tabular}
\end{table}

\begin{table}[t]
\centering
\scriptsize
\setlength{\tabcolsep}{1.5pt}
\caption{Capacity expansion: paired cost differences (bn\$/yr; negative: the first method is cheaper) at each radius pair of Table~\ref{tab:cep-pareto}: the median over records (standard error) of each record's median over seeds. $^{*}$/$^{**}$: two-sided Wilcoxon signed-rank $p < 0.05$/$0.01$ over records. With five records ($k = 1$) or blocks ($k = 5$), the smallest attainable $p$ is $0.0625$.}
\label{tab:cep-paired}
\begin{tabular}{cccccccccc}
\toprule
 & & \multicolumn{4}{c}{In-distribution} & \multicolumn{4}{c}{Out-of-distribution} \\
\cmidrule(lr){3-6}\cmidrule(lr){7-10}
$\rho_{\mathrm{ref}}$ & $\rho$ & \shortstack{KL-Gen\\$-$ KL-Rec} & \shortstack{KL-Gen\\$-$ KL-OT} & \shortstack{KL-Gen-P\\$-$ KL-Gen} & \shortstack{KL-Gen\\$-$ KL-Gen\\(copula)} & \shortstack{KL-Gen\\$-$ KL-Rec} & \shortstack{KL-Gen\\$-$ KL-OT} & \shortstack{KL-Gen-P\\$-$ KL-Gen} & \shortstack{KL-Gen\\$-$ KL-Gen\\(copula)} \\
\midrule
\multicolumn{10}{l}{\textit{$k = \frac12$ (10 record years)}} \\
0.2 & 0 & $-2.92$ (1.02)$^{*}$ & $+0.14$ (0.30) & $-0.07$ (0.12) & $+0.30$ (0.15) & $-22.21$ (4.64)$^{**}$ & $-2.65$ (2.22)$^{*}$ & $-0.27$ (0.75) & $-1.44$ (0.87) \\
 & 0.05 & $-1.72$ (1.06) & $+0.28$ (0.30) & $-0.05$ (0.14) & $+0.19$ (0.16) & $-18.29$ (5.13)$^{**}$ & $-2.41$ (2.85) & $-0.07$ (0.30) & $-0.55$ (0.90) \\
 & 0.2 & $-1.19$ (0.78)$^{*}$ & $+0.31$ (0.15) & $+0.20$ (0.15) & $-0.06$ (0.25) & $-14.95$ (3.86)$^{**}$ & $-2.28$ (1.48)$^{**}$ & $-0.10$ (0.58) & $+0.05$ (0.58) \\
\cmidrule(lr){1-10}
0.4 & 0 & $-1.93$ (1.14)$^{*}$ & $+0.12$ (0.45) & $+0.04$ (0.18) & $+0.16$ (0.15) & $-17.66$ (4.45)$^{**}$ & $-6.51$ (2.96)$^{*}$ & $-0.13$ (1.36) & $-0.03$ (1.10) \\
 & 0.05 & $-0.46$ (1.18) & $+0.16$ (0.24) & $+0.10$ (0.14) & $-0.02$ (0.15) & $-12.71$ (5.57)$^{**}$ & $-4.09$ (1.46)$^{**}$ & $+0.01$ (0.66) & $+0.30$ (1.33) \\
\midrule
\multicolumn{10}{l}{\textit{$k = 1$ (5 record years)}} \\
0.2 & 0 & $-0.47$ (0.34) & $+0.30$ (0.50) & $-0.16$ (0.05) & $+0.19$ (0.04) & $-14.51$ (1.34) & $-1.93$ (1.93) & $-0.03$ (1.47) & $-0.16$ (1.33) \\
 & 0.05 & $+0.45$ (0.26) & $+0.45$ (0.23) & $-0.19$ (0.29) & $+0.24$ (0.20) & $-8.42$ (2.78) & $-1.46$ (3.47) & $-0.65$ (3.39) & $+0.16$ (2.22) \\
 & 0.2 & $+0.16$ (0.33) & $+0.17$ (0.12) & $-0.00$ (0.28) & $+0.06$ (0.13) & $-9.23$ (1.25) & $-0.46$ (1.87) & $-0.02$ (1.59) & $-0.03$ (1.25) \\
\cmidrule(lr){1-10}
0.4 & 0 & $-0.43$ (0.33) & $+0.14$ (0.25) & $+0.33$ (0.24) & $-0.07$ (0.21) & $-12.63$ (2.71) & $-1.33$ (1.18) & $-0.29$ (0.66) & $+0.00$ (0.57) \\
 & 0.05 & $-0.23$ (0.59) & $+0.42$ (0.18) & $+0.20$ (0.08) & $+0.22$ (0.18) & $-14.31$ (1.87) & $-0.39$ (0.85) & $-0.11$ (0.30) & $-0.00$ (0.43) \\
\midrule
\multicolumn{10}{l}{\textit{$k = 5$ (5 record blocks)}} \\
0.2 & 0 & $+0.25$ (0.25) & $-0.11$ (0.09) & $+0.18$ (0.04) & $+0.07$ (0.09) & $-8.90$ (2.50) & $-0.07$ (0.34) & $-0.14$ (0.89) & $-0.79$ (0.82) \\
 & 0.05 & $-0.07$ (0.31) & $+0.09$ (0.08) & $+0.01$ (0.16) & $+0.31$ (0.26) & $-13.80$ (2.06) & $-0.39$ (0.81) & $-0.28$ (0.13) & $-0.84$ (0.35) \\
 & 0.2 & $+0.18$ (0.17) & $+0.38$ (0.36) & $-0.08$ (0.22) & $-0.04$ (0.19) & $-10.19$ (1.77) & $-0.33$ (0.58) & $-0.35$ (0.17) & $+0.17$ (0.85) \\
\cmidrule(lr){1-10}
0.4 & 0 & $+0.57$ (0.34) & $+0.33$ (0.10) & $-0.13$ (0.08) & $+0.26$ (0.07) & $-6.50$ (1.87) & $-0.76$ (0.50) & $+0.37$ (0.31) & $-0.15$ (0.67) \\
 & 0.05 & $+0.73$ (0.15) & $+0.24$ (0.19) & $+0.31$ (0.16) & $+0.36$ (0.15) & $-3.79$ (1.97) & $-0.18$ (0.74) & $-0.08$ (0.11) & $-0.31$ (0.15) \\
\bottomrule
\end{tabular}
\end{table}

\end{document}